\documentclass[aps,prd,amsmath,amssymb,showpacs]{revtex4}
\usepackage{amssymb}
\usepackage{mathbbol}
\usepackage{amsfonts}
\usepackage{mathrsfs}
\usepackage{epsfig,bm,dcolumn}
\usepackage{graphicx}
\usepackage{color}
\usepackage{amsmath}
\usepackage{dcolumn}
\usepackage{overpic}
\usepackage{hyperref}
\usepackage{slashed}

\begin{document}
%\begin{CJK}{GBK}{song}

\title{Complete and Consistent Chiral Transport from Wigner Function Formalism}
\author{Anping Huang$^a$}
\author{Shuzhe Shi$^b$}
\author{Yin Jiang$^{c}$}
\author{Jinfeng Liao$^{b}$}\email{liaoji@indiana.edu}
\author{Pengfei Zhuang$^a$}\email{zhuangpf@mail.tsinghua.edu.cn}
\address{$^a$Physics Department, Tsinghua University, Beijing 100084, China.\\
	$^b$Physics Department and Center for Exploration of Energy and Matter,	Indiana University, 2401 N Milo B. Sampson Lane, Bloomington, IN 47408, USA.\\
	$^c$School of Physics and Nuclear Energy Engineering, Beihang University, Beijing 100191, China.}
%\email{liaoji@indiana.edu}
%\email{zhuangpf@mail.tsinghua.edu.cn}

\date{\today}

\begin{abstract} 
Recently there has been significant interest in understanding the macroscopic quantum transport  in  a many-body system of chiral fermions. A natural  framework for describing such a system which is generally  out of equilibrium, is the transport equation for its phase space distribution function. 
In this paper, we obtain a complete solution of the covariant chiral transport for massless fermions, by starting from the general Wigner function formalism and carrying out a complete and consistent semiclassical expansion up to $\hat{\mathbf{O}}(\hbar)$ order. In particular, we clarify certain subtle and confusing issues surrounding the Lorentz non-invariance and frame dependence associated with the 3D chiral kinetic theory.  We prove that such frame dependence is uniquely and completely fixed by an unambiguous definition of the $\hat{\mathbf{O}}(\hbar)$ correction to the distribution function in each reference frame. 
\end{abstract}

\maketitle

%============================================================================
\section{Introduction}

The many-body physics of massless fermions has attracted significant interest in a wide range of communities, from condensed matter physics to high energy heavy ion collisions. In particular, the microscopic quantum anomaly of such chiral fermions can induce highly nontrivial macroscopic transport phenomena, such as the notable example of Chiral Magnetic Effect~\cite{Kharzeev:2004ey, Kharzeev:2007tn, Kharzeev:2007jp, Fukushima:2008xe} as well as the chiral vortical effect (CVE)~\cite{Son:2009tf, Kharzeev:2010gr, Jiang:2015cva}. These effects have been extensively studied using various many-body theoretical tools~\cite{Erdmenger:2008rm,Banerjee:2008th, Torabian:2009qk,Kalaydzhyan:2011vx,Son:2009tf,Pu:2010as,Kharzeev:2011ds,Metlitski:2005pr,Newman:2005as,Charbonneau:2009ax,Lublinsky:2009wr,Asakawa:2010bu,Landsteiner:2011cp}.  Enthusiastic efforts have also been made to experimentally measure such anomalous chiral transport effects, both in the so-called Dirac or Weyl semimetals and in the so-called quark-gluon plasma created via heavy ion collisions. For  reviews  on recent developments, see e.g. \cite{Kharzeev:2015znc,Kharzeev:2015kna,Liao:2014ava,Hattori:2016emy,Burkov:2015hba}.

An important aspect of the many-body theory for  anomalous chiral transport is to describe the out-of-equilibrium situation. The natural framework is the kinetic theory based on transport equations for the phase space distribution function of such a system. Different from usual classical kinetic theory~\cite{pitaevskii2017course}, a proper description of the chiral fermions must account for intrinsic quantum and relativistic effects.  A lot of progress has been achieved lately to develop such a chiral kinetic theory, see e.g.~\cite{Son:2012wh,Son:2012zy,Stephanov:2012ki,Chen:2014cla,Chen:2015gta,Kharzeev:2016sut,Chen:2012ca,Gao:2012ix,Gao:2017gfq,Hidaka:2016yjf,Hidaka:2017auj,Mueller:2017lzw,Mueller:2017arw,Gorbar:2017cwv,Wu:2016dam}.  There also exist a lot of phenomenological interests and attempts to study anomalous chiral transport in the out-of-equilibrium setting~\cite{Mace:2016svc,Mace:2016shq,Mueller:2016ven,Fukushima:2015tza,Ebihara:2017suq,Sun:2016nig,Sun:2016mvh,Huang:2017tsq,Jiang:2016wve,Shi:2017cpu}. The transport theory of chiral fermions, however, bears unusual subtlety and poses a number of challenges, particularly related to Lorentz invariance and frame dependence. A resolution was developed in the 3D formulation of chiral kinetic theory \cite{Chen:2014cla,Chen:2015gta,Hidaka:2016yjf}, but the origin of such issues remains  cloudy. It is highly desirable to develop a transport theory of chiral fermions in a completely covariant fashion and to identify the precise reason of these complications.

A natural approach is to derive the quantum transport equation for chiral fermions in the well-known Wigner function formalism by a systematic semiclassical expansion in terms of $\hbar$~\cite{Vasak:1987um,Zhuang:1995pd,Zhuang:1995jb,Zhuang:1998kv,Ochs:1998qj,Guo:2017dzf}. We shall adopt this approach in the present paper. We will systematically derive the chiral transport equations for a general out-of-equilibrium system of  collision-less massless fermions, under external electromagnetic fields that are generally space-time dependent.  The start point is the Wigner function and the kinetic equation for Winger function and its 16   components, such as the vector $\mathscr{V}_{\mu}$, axial vector $\mathscr{A}_{\mu}$, scarlar $\mathscr{F}$, pseudoscalar $\mathscr{P}$, antisymmetry tensor $\mathscr{L}^{\mu\nu}$. These 16   components would be decoupled for chiral fermion system.  We will focus on the set of equations for vector $\mathscr{V}^{\mu}$ and axial vector $\mathscr{A}^{\mu}$ components. By carrying out the semi-classical expansion for all the operators and functions, one can then derive   a covariant set of chiral transport equations to $\hbar$ order. In particular, this detailed derivation will allow    a clear understanding, within a totally covariant framework, of the origin for the rather confusing Lorentz invariance and frame dependence issues as well as the emergence of the Berry phase, Berry curvature and anomalous terms in the 3D formulation of chiral kinetic theory. In fact, we will prove that such frame dependence is uniquely and completely fixed by an unambiguous definition of the $\hat{\mathbf{O}}(\hbar)$ correction to the distribution function in each reference frame.

The paper is organized as follows. In Sec. \ref{sec:1}, we briefly review
the Wigner function formalism and the kinetic equations for the 16  components of Wigner function. In Sec. \ref{sec:2}, these equations are decoupled and decompose to two set of equations for the massless case and we focus on the semi-classical expansion for the chiral currents. With the obtained constraint equations we construct the most general solutions and discuss the frame dependence issue. 
In Sec. \ref{sec:3} we present the covariant chiral transport equations as well as their 3D formulation. Finally we conclude the paper in In Sec. \ref{sec:4}. An appendix is also included to particularly prove in great technical details the completeness and uniqueness of the found $\hat{\mathbf{O}}(\hbar)$ solution to the constraint equations which is crucial for understanding the frame dependence issue.

%============================================================================
\section{The quantum kinetic equations in the Wigner function formalism}\label{sec:1}

The bridge connecting quantum field theory to relativistic kinetic theory is the Wigner function\cite{DeGroot:1980dk,Vasak:1987um}.
%, which is the ensemble average of Wigner operator.
For the Dirac field $\psi$ with charge $Q$, the general gauge invariant Wigner operator is defined as
\begin{align}
&\hat{W}_{\alpha\beta}(x,p)=\int\frac{d^{4}y}{(2\pi)^4}e^{-\frac{i}{\hbar}p\cdot y} \bar{\psi}_{\beta}(x_{+})U(x_{+},x_{-})\psi_{\alpha}(x_{-}),
\end{align}
where $\alpha$ and $\beta$ are spinor indices. Also, the gauge link $U$ between $x_{\pm}=x\pm y/2$ is introduced to ensure the gauge invariance of the Wigner operator. It's defined as
\begin{eqnarray}
U(x_{+},x_{-})=\mathcal{P}e^{-\frac{iQ}{\hbar} y^{\mu}\int^{1}_{0}ds A_{\mu}(x-\frac{y}{2}+sy)},
\end{eqnarray}
where the path-ordering operator $\mathcal{P}$ can be dropped for abelian $A^\mu$ fields. In this work, we keep the  Planck constant $\hbar$ in various places  to show quantum effect explicitly.

%Here we have reinstated $\hbar$ into Wigner operator to show the quantum charaCKEr.
%
%Obviously, the Wigner operator transforms like an ordinary $\gamma$ matrix($\gamma^{\mu\dagger}=\gamma^{0}\gamma^{\mu}\gamma^{0}$),
%\begin{align}\label{eq:070}
%&\hat{W}^{\dagger}(x,p)=\gamma^{0}\hat{W}(x,p)\gamma^{0}.
%\end{align}
%This property ensures the physical quantities, such as vector currents  $\mathrm{tr}(\gamma^{\mu}W)$ and axial vector currents $\mathrm{tr}(\gamma^{\mu}\gamma^{5}W)$,  are real.

%With the ensemble averaging $\langle \cdot \rangle$ the Winger function is defined as
Then one can construct the Winger function, as the expectation value of the Wigner operator
\begin{eqnarray}
W_{\alpha\beta}(x,p)=\left< \hat{W}_{\alpha\beta}(x,p) \right>,
\end{eqnarray}
where $\langle \cdot \cdot \cdot \rangle$ means the expectation over a given quantum state, or the average over an ensemble of quantum states.

In this work, we consider a collisionless system in a background electromagnetic field $A^{\mu}$.
In this case the Wigner function satisfies the quantum kinetic equation \cite{Vasak:1987um}
\begin{eqnarray}\label{eq:049}
\left(\slashed{\mathbf{K}}-m\right)W(x,p)=0 \,\, ,
\end{eqnarray}
where $\slashed{\mathbf{K}}=\gamma^{\mu} \mathbf{K}_{\mu}$, $\mathbf{K}_{\mu}=\pi_{\mu}+\frac{1}{2}i\hbar\triangledown_{\mu}$, and
\begin{eqnarray}
\pi^{\mu}&=&p^{\mu}-\frac{1}{2}Q\hbar \, j_{1}\left(\frac{1}{2}\hbar \triangle \right) \, F^{\mu\nu}  \partial^{p}_{\nu},\\
\triangledown^{\mu}&=&\partial^{\mu}-Q \, j_{0}\left(\frac{1}{2}\hbar \triangle \right) \, F^{\mu\nu} \partial^{p}_{\nu}.
\end{eqnarray}
Note that in the triangle operator $\triangle=\partial_{x}\cdot\partial_{p}$, $\partial_{x}$ acts only on electromagnetic tensor $F_{\mu\nu}=\partial^{\mu}A^{\nu}-\partial^{\nu}A^{\mu}$, while $\partial_{p}$ acts only on $W(x,p)$. In addition, $j_{0}(x)=x^{-1}\sin(x)$ and $j_{1}(x)=x^{-2}\sin(x)-x^{-1}\cos(x)$ are the spherical Bessel functions which are generated by the y-integrations. 
In general combining with the Maxwell equation, the quantum kinetic equation of Wigner function Eq.(\ref{eq:049}) is equivalent to the QED field theory.

In order to connect Eq.(\ref{eq:049}) with  kinetic theory,
one needs to obtain explicitly the equations of all elements of the Wigner function, which is a $4\times 4$ matrix.
In order to do that, one can decompose the $W(x,p)$ in terms of the 16  generators of the Clifford algebra,
choosing the convention basis as follows:
\begin{eqnarray}
&&\Gamma^a=I,~\gamma^\mu,~i\gamma^5,~\gamma^\mu\gamma^5,~\sigma^{\mu\nu},\nonumber\\
&&\Gamma_a=I,~\gamma_\mu,~-i\gamma_5,~\gamma_5\gamma_\mu,~\sigma_{\mu\nu}.
%\nonumber\\&&F=\frac{1}{4}\Sigma_{a}f_a\Gamma^a,~~~f_a=\mathrm{tr}\left(\Gamma_a F\right).
\end{eqnarray}
In this basis, the Wigner function is expanded as
\begin{equation}\label{eq:071}
W=\frac{1}{4}\left(\mathscr{F}+i\gamma^5\mathscr{P}+\gamma^\mu\mathscr{V}_{\mu}+\gamma^\mu\gamma^5\mathscr{A}_{\mu}+\frac{1}{2}\sigma^{\mu\nu}\mathscr{L}_{\mu\nu}\right),
\end{equation}
where these sixteen components are given by
\begin{eqnarray}
\mathscr{F}(x,p) &=& \mathrm{tr}W(x,p),\nonumber\\
\mathscr{P}(x,p) &=& -i\mathrm{tr}[\gamma_5 W(x,p)],\nonumber\\
\mathscr{V}_{\mu}(x,p) &=& \mathrm{tr}[\gamma_\mu W(x,p)],\nonumber\\
\mathscr{A}_{\mu}(x,p) &=& \mathrm{tr}[\gamma_5 \gamma_\mu W(x,p)],\nonumber\\
\mathscr{L}_{\mu\nu}(x,p) &=& \mathrm{tr}[\sigma_{\mu\nu} W(x,p)]=-\mathscr{L}_{\nu\mu}(x,p).
\end{eqnarray}
Noting that the Wigner function satisfies hermiticity relations $W^{\dagger}(x,p)=\gamma^{0} W(x,p)\gamma^{0}$ in the same way as the $\Gamma_a$'s ($\Gamma^{\dagger}_{a}=\gamma^{0}\Gamma_{a}\gamma^{0}$),
all these 16 components are real, and they behave as scalar, pseudo-scalar, vector, axial vector and antisymmetric tensor, respectively, under Lorentz transformation. Each of these sixteen components is connected with a corresponding physical quantity\cite{BialynickiBirula:1991tx,Zhuang:1995pd}. Explicitly speaking the vector $\mathscr{V}_{\mu}$ and axial vector $\mathscr{A}_{\mu}$ can be used to construct the current density $J^{\mu}$, axial current density $J^{\mu}_{5}$ and energy-momentum tensor $T^{\mu\nu}$,
\begin{align}\begin{split}
&J^{\mu}(x)=\left<\bar{\psi}(x)\gamma^{\mu}\psi(x)\right>=\int d^4 p~\mathrm{tr}\left( \gamma^{\mu}W(x,p)\right)=\int d^4 p \mathscr{V}^{\mu}(x,p),\\
&J^{\mu}_{5}(x)=\left<\bar{\psi}(x)\gamma^{\mu}\gamma^{5}\psi(x)\right>=-\int d^4 p~\mathrm{tr}\left( \gamma^{5}\gamma^{\mu}W(x,p)\right)=-\int d^4 p \mathscr{A}^{\mu}(x,p),\\
&T^{\mu\nu}(x)=\frac{-i}{2}\left<\bar{\psi}(x)\left[\gamma^{\mu}D^{\nu +}-\gamma^{\nu}D^{\mu}\right]\psi(x)\right>=\int d^4 p~p^{\nu}\mathrm{tr}\left( \gamma^{\mu}W(x,p)\right)=\int d^4 p ~p^{\nu}\mathscr{V}^{\mu}(x,p).
\end{split}\end{align}

Now, we can derive the kinetic equations for these 16 coefficients explicitly. Substituting the decomposed Wigner function Eq.(\ref{eq:071}) into  Eq.(\ref{eq:049}), one obtains:  
\begin{eqnarray}\label{eq:072}
0&=&\left(\gamma^{\mu}\mathbf{K}_{\mu}\mathscr{F}+i\gamma^{\mu}\gamma^{5}\mathbf{K}_{\mu}\mathscr{P}+\gamma^{\mu}\gamma^{\nu}\mathbf{K}_{\mu}\mathscr{V}_{\nu}+\gamma^{\mu}\gamma^{\nu}\gamma^{5}\mathbf{K}_{\mu}\mathscr{A}_{\nu}+\frac{1}{2}\gamma^{\mu}\sigma^{\nu\sigma}\mathbf{K}_{\mu}\mathscr{L}_{\nu\sigma}\right)\nonumber\\
&&-m\left(\mathscr{F}+i\gamma^5\mathscr{P}+\gamma^\mu\mathscr{V}_{\mu}+\gamma^\mu\gamma^5\mathscr{A}_{\mu}+\frac{1}{2}\sigma^{\mu\nu}\mathscr{L}_{\mu\nu}\right).
\end{eqnarray}
Next we will use the following properties of the $\gamma$ matrices (with the metric convention $g^{\mu\nu}=\mathrm{diag}(1,-1,-1,-1)$, and the Levi-Civita anti-symmetric tensor $\epsilon^{0123}=-\epsilon_{0123}=1$),  
\begin{eqnarray*}
&&\{\gamma^{\mu},\gamma^{5}\}=0,\qquad\{\gamma^{\mu},\gamma^{\nu}\}=2g^{\mu\nu},\qquad\gamma^{\mu}\gamma^{\nu}=g^{\mu\nu}-i\sigma^{\mu\nu},\\
&&\sigma^{\mu\nu}\gamma^{5}=\frac{i}{2}\epsilon^{\mu\nu\sigma\rho}\sigma_{\sigma\rho},\qquad\gamma^{\sigma}\sigma^{\mu\nu}=g^{\sigma\mu}\gamma^\nu-g^{\sigma\nu}\gamma^\mu+i\epsilon^{\mu\nu\sigma\rho}\gamma_\rho\gamma^5 \, ,
\end{eqnarray*}
 to cast  terms with multiple $\gamma$ matrices into $\Gamma_a$ basis: 
\begin{eqnarray*}
&&\gamma^{\mu}\gamma^{\nu}\mathbf{K}_{\mu}\mathscr{V}_{\nu}=\mathbf{K}_{\mu}\mathscr{V}^{\mu}-\frac{i}{2}\sigma^{\mu\nu}\left(\mathbf{K}_{\mu}\mathscr{V}_{\nu}-\mathbf{K}_{\nu}\mathscr{V}_{\mu}\right),\\
&&\gamma^{\mu}\gamma^{\nu}\gamma^{5}\mathbf{K}_{\mu}\mathscr{A}_{\nu}=-i\left( i\gamma^{5}\right)\mathbf{K}_{\mu}\mathscr{A}^{\mu}+\frac{1}{2}\epsilon_{\mu\nu\sigma\rho}\sigma^{\mu\nu}\mathbf{K}^{\sigma}\mathscr{A}^{\rho},\\
&&\gamma^{\mu}\sigma^{\nu\sigma}\mathbf{K}_{\mu}\mathscr{L}_{\nu\sigma}= -2i\gamma^\mu\mathbf{K}^{\nu}\mathscr{L}_{\mu\nu}+\epsilon_{\mu\nu\sigma\rho}\gamma^{\mu}\gamma^5\mathbf{K}^{\nu}\mathscr{L}^{\sigma\rho} \, .
\end{eqnarray*}
These relations allow us to  simplify Eq.(\ref{eq:072}) as
\begin{eqnarray}
&&I\Big(\mathbf{K}_{\mu}\mathscr{V}^{\mu}-m\mathscr{F}\Big)
+i\gamma^5\Big(-i\mathbf{K}_{\mu}\mathscr{A}^{\nu}-m\mathscr{P}\Big)
+\gamma^{\mu}\Big(\mathbf{K}_{\mu}\mathscr{F}-i\mathbf{K}^{\nu}\mathscr{L}_{\mu\nu}-m\mathscr{V}_{\mu}\Big)\nonumber\\
&&+\gamma^{\mu}\gamma^5\Big(i\mathbf{K}_{\mu}\mathscr{P}+\frac{1}{2}\epsilon_{\mu\nu\sigma\rho}\mathbf{K}^{\nu}\mathscr{L}^{\sigma\rho}-m\mathscr{A}_{\mu}\Big)
+\frac{1}{2}\sigma^{\mu\nu}\Big(-i(\mathbf{K}_{\mu}\mathscr{V}_{\nu}-\mathbf{K}_{\nu}\mathscr{V}_{\mu})+\epsilon_{\mu\nu\sigma\rho}\mathbf{K}^{\sigma}\mathscr{A}^{\rho}-m\mathscr{L}_{\mu\nu}\Big)=0.
\end{eqnarray}

From the orthogonality of $\{\Gamma_a\}$ basis, i.e. $\mathrm{tr}(\Gamma_a\Gamma_b)=4\delta_{ab}$, one can prove that all ``elements'' of the above ``matrix'' should be zero, {\it i.e.}
\begin{eqnarray}
0&=&\label{eq:001}\mathbf{K}_{\mu}\mathscr{V}^{\mu}-m\mathscr{F},\\
0&=&\label{eq:002}i\mathbf{K}_{\mu}\mathscr{A}^{\mu}+m\mathscr{P},\\
0&=&\label{eq:003}\mathbf{K}_{\mu}\mathscr{F}-i\mathbf{K}^{\nu}\mathscr{L}_{\mu\nu}-m\mathscr{V}_{\mu},\\
0&=&\label{eq:004}i\mathbf{K}_{\mu}\mathscr{P}+\frac{1}{2}\epsilon_{\mu\nu\sigma\rho}\mathbf{K}^{\nu}\mathscr{L}^{\sigma\rho}-m\mathscr{A}_{\mu},\\
0&=&\label{eq:005}i\left(\mathbf{K}_{\mu}\mathscr{V}_{\nu}-\mathbf{K}_{\nu}\mathscr{V}_{\mu}\right)-\epsilon_{\mu\nu\sigma\rho}\mathbf{K}^{\sigma}\mathscr{A}^{\rho}+m\mathscr{L}_{\mu\nu}.
\end{eqnarray}
Furthermore, as $\mathbf{K}^{\mu}=\pi^{\mu}+\frac{1}{2}i\hbar\triangledown^{\mu}$ is complex while all components of the Wigner function are real,
one could further separate the above equations with the real and imaginary parts. The real parts give
\begin{eqnarray}
m\mathscr{F}&=&\pi^{\mu}\mathscr{V}_{\mu},\label{eq:008}\\
2m\mathscr{P}&=&\hbar\triangledown^{\mu}\mathscr{A}_{\mu},\label{eq:009}\\
m\mathscr{V}_{\mu}&=&\pi_{\mu}\mathscr{F}+\frac{1}{2}\hbar\triangledown^{\nu}\mathscr{L}_{\mu\nu},\label{eq:010}\\
2m\mathscr{A}_{\mu}&=&-\hbar\triangledown_{\mu}\mathscr{P}+\epsilon_{\mu\nu\sigma\rho}\pi^{\nu}\mathscr{L}^{\sigma\rho},\label{eq:011} \\
m\mathscr{L}_{\mu\nu}&=&\frac{1}{2}\hbar\left(\triangledown_{\mu}\mathscr{V}_{\nu}-\triangledown_{\nu}\mathscr{V}_{\mu}\right)+\epsilon_{\mu\nu\sigma\rho}\pi^{\sigma}\mathscr{A}^{\rho},\label{eq:012}
\end{eqnarray}
while the imaginary parts lead to
\begin{eqnarray}
&&\hbar\triangledown^{\mu}\mathscr{V}_{\mu}=0,\label{eq:013}\\
&&\pi^{\mu}\mathscr{A}_{\mu}=0,\label{eq:014}\\
&&\frac{1}{2}\hbar\triangledown_{\mu}\mathscr{F}=\pi^{\nu}\mathscr{L}_{\mu\nu},\label{eq:015}\\
&&\pi_{\mu}\mathscr{P}=-\frac{1}{4}\hbar\epsilon_{\mu\nu\sigma\rho}\triangledown^{\nu}\mathscr{L}^{\sigma\rho},\label{eq:016}\\
&&\pi_{\mu}\mathscr{V}_{\nu}-\pi_{\nu}\mathscr{V}_{\mu}=\frac{1}{2}\hbar\epsilon_{\mu\nu\sigma\rho}\triangledown^{\sigma}\mathscr{A}^{\rho}.\label{eq:017}
\end{eqnarray}
The above results are the complete quantum kinetic equations~\cite{Vasak:1987um,Zhuang:1995pd,Zhuang:1995jb,Zhuang:1998kv,Ochs:1998qj}, as shown in Eq.(\ref{eq:008} - \ref{eq:017}), in terms of the 16 components of the Wigner function which are  coupled with each other. In the next section, we will focus on the massless case to further simplify the kinetic equations.

%============================================================================
\section{Chiral Transport Equations and the General Solutions}\label{sec:2}
%In this section, we consider the massless system, and try to systematically derive the covariant chiral kinetic equation for chiral system from the kinetic equations of Wigner function. There are some interesting charaCKErs for the chiral system, such as chiral magnetic effect and triangle anomalous effect, which are absent in the classical level. We expect that the new covariant chiral kinetic equation can also consistently contain these anomalous quantum effects.

In this section, we consider a system of chiral fermions with $m=0$. In this case, the quantum kinetic equations in Eq.(\ref{eq:008} - \ref{eq:017}) get partially decoupled.
One can see explicitly that they are separated into two  groups:
a set of equations describing the evolution of scalar $\mathscr{F}$, pseudoscalar $\mathscr{P}$ and antisymmetry tensor $\mathscr{L}^{\mu\nu}$ components 
\begin{align}\begin{split}\label{eq:059}
&\pi_{\mu}\mathscr{F}+\frac{1}{2}\hbar\triangledown^{\nu}\mathscr{L}_{\mu\nu}=0,\\
&\frac{1}{2}\hbar\triangledown_{\mu}\mathscr{F}-\pi^{\nu}\mathscr{L}_{\mu\nu}=0,\\
&-\hbar\triangledown_{\mu}\mathscr{P}+\epsilon_{\mu\nu\rho\sigma}\pi^{\nu}\mathscr{L}^{\rho\sigma}=0, \\
&\pi_{\mu}\mathscr{P}+\frac{1}{4}\hbar\epsilon_{\mu\nu\rho\sigma}\triangledown^{\nu}\mathscr{L}^{\rho\sigma}=0,\\
\end{split}\end{align}
and another set for vector $\mathscr{V}_{\mu}$ and axial vector $\mathscr{A}_{\mu}$ components: 
\begin{align}\begin{split}\label{eq:051}
&\pi^{\mu}\mathscr{V}_{\mu}=0,\qquad\pi^{\mu}\mathscr{A}_{\mu}=0,\\
&\hbar\triangledown^{\mu}\mathscr{V}_{\mu}=0,\qquad\hbar\triangledown^{\mu}\mathscr{A}_{\mu}=0, \\
&\hbar\epsilon_{\mu\nu\rho\sigma}\triangledown^{\rho}\mathscr{V}^{\sigma}=2(\pi_{\mu}\mathscr{A}_{\nu}-\pi_{\nu}\mathscr{A}_{\mu}), \\
&\hbar\epsilon_{\mu\nu\rho\sigma}\triangledown^{\rho}\mathscr{A}^{\sigma}=2(\pi_{\mu}\mathscr{V}_{\nu}-\pi_{\nu}\mathscr{V}_{\mu}).
\end{split}\end{align}
%The last line of above equations is from Eq.(\ref{eq:012}) with the relation $\epsilon_{\tau\lambda\mu\nu}\epsilon^{\mu\nu\sigma\beta}=-2\left(\delta^{\sigma}_{\tau}\delta^{\beta}_{\lambda}-\delta^{\sigma}_{\lambda}\delta^{\beta}_{\tau}\right)$.

Noting the specific patterns of vector (scalar) and axial-vector (pseudo-scalar) terms, one could further simplify the above two sets of equations by introducing the ``chiral basis''\cite{Ochs:1998qj, Gao:2017gfq} via
\begin{align}\begin{split}
\mathscr{T}_{\chi}&=\frac{1}{2}(\mathscr{F}+\chi\mathscr{P}),\\
\mathscr{S}^{\mu\nu}_{\chi}&=\frac{1}{2}\left(\mathscr{L}^{\mu\nu}+\chi\frac{1}{2}\epsilon^{\mu\nu\sigma\rho}\mathscr{L}_{\sigma\rho} \right),\\
\mathscr{J}^{\mu}_{\chi}&=\frac{1}{2}(\mathscr{V}^{\mu}-\chi\mathscr{A}^{\mu}),
\end{split}\end{align}
where $\chi=\pm1$ corresponds to the chirality of massless fermion. In such chiral basis, Eq.(\ref{eq:059}) can be further decomposed, in which the right-handed(RH) and left-handed(LH) components get decoupled:
\begin{eqnarray}
&&\pi_{\mu}\mathscr{T}_\chi+\frac{1}{2}\hbar\triangledown^{\nu}\mathscr{S}_{\mu\nu}^\chi=0,\\
&&\pi_{\mu}\mathscr{S}^{\mu\nu}_\chi+\frac{1}{2}\hbar\triangledown^{\nu}\mathscr{T}_\chi=0 \, .
\end{eqnarray}
Similarly  Eq.(\ref{eq:051}) can be recast into RH and LH sectors:  
\begin{eqnarray}
&&\hbar\epsilon_{\mu\nu\rho\sigma}\triangledown^{\rho}\mathscr{J}^{\sigma}_\chi=-2\chi(\pi_{\mu}\mathscr{J}_{\nu}^\chi-\pi_{\nu}\mathscr{J}_{\mu}^\chi),
\label{eq.CKE.1}\\
&&\pi^{\mu}\mathscr{J}_{\mu}^\chi=0, \label{eq.CKE.2}\\
&&\triangledown^{\mu}\mathscr{J}_{\mu}^\chi=0 \label{eq.CKE.3}.
\end{eqnarray}
The decoupling of the RH and LH components in these equations reflects a basic property of massless fermions: for the massless Dirac fermions, the RH and LH sectors can be completely separated in the Lagrangian. 

As the main purpose of this paper is to study the chiral transport effects,  we will focus on the  equations for the chiral components $\mathscr{J}^{\mu}_{\chi}$, namely the Eqs. (\ref{eq.CKE.1}-\ref{eq.CKE.3}) in the following.  
We note in passing that the chiral components $\mathscr{J}^{\mu}_{\chi}$ can be directly related to the  physical chiral currents: 
\begin{eqnarray}\label{eq:067}
& J^{\mu}_{\chi}=\left<\bar{\psi}_{\chi}\gamma^{\mu}\gamma^{5}\psi_{\chi}\right>=\int d^{4}p \mathscr{J}^{\mu}_{\chi}=\frac{1}{2}\left(J^{\mu}+\chi J^{\mu}_{5}\right).
\end{eqnarray}
Here $\psi_{\chi}=P_{\chi}\psi$ and  $\bar{\psi}_{\chi}=\bar{\psi}P_{-\chi}$, with $P_{\chi}=(1+\chi\gamma^{5})/2$ being the chirality projection operators.

%============================================================================
\subsection{Semi-classical expansion}\label{sec.hexpansion}

We now derive the chiral kinetic equation, by starting from Eqs.(\ref{eq.CKE.1}-\ref{eq.CKE.3}) and utilizing the semi-classical expansion method~\cite{Vasak:1987um}. In order to do this, one needs to expand both operators and Wigner function components in the evolution equations order by order in terms of $\hbar$.  
First of all, let's expand the operators $\pi^{\mu}$ and $\triangledown^{\mu}$ in powers of $\hbar$, by using the Taylor expansion of the  spherical Bessel function $j_{0}$ and $j_{1}$  in terms of $\frac{1}{2}\hbar\triangle$, with $j_{0}(x)=1-x^2/6+\mathbf{O}(x^4)$ and  $j_{1}(x)=x/3-x^3/30+\mathbf{O}(x^5)$: 
\begin{eqnarray}\begin{split}\label{eq:074}
&\pi^{\mu}=p^{\mu}-\frac{1}{2}Q\hbar j_{1}\left(\frac{1}{2}\hbar \triangle \right)F^{\mu\nu} \partial^{p}_{\nu}=p^{\mu}-\frac{1}{12}Q\hbar^2\triangle F^{\mu\nu}\partial^{p}_{\nu}+\mathbf{O}(\hbar^4),\\
&\triangledown^{\mu}=\partial^{\mu}-Q j_{0}\left(\frac{1}{2}\hbar \triangle \right)F^{\mu\nu} \partial^{p}_{\nu}=\partial^{\mu}-QF^{\mu\nu}\partial^{p}_{\nu}+\frac{1}{24}Q\hbar^2\triangle^2 F^{\mu\nu}\partial^{p}_{\nu}+\mathbf{O}(\hbar^4).
\end{split}\end{eqnarray}
The truncation  of this expansion series would be justified when $\frac{1}{2}\hbar |\partial_{x}F^{\mu\nu} \cdot \partial_{p}W(x,p)|\ll |F^{\mu\nu}W(x,p)|$. In other words, the electromagnetic field $F^{\mu\nu}$ and Wigner function $W(x,p)$ should vary smoothly enough in coordinate space $x$ and momentum space $p$, respectively\cite{Vasak:1987um,Yagi:2005yb}. It should be emphasized that, starting from here through the rest of this paper, we will use the notation $\triangledown^{\mu}$ to represent its zeroth-order truncation, i.e. $\triangledown^{\mu}\to \partial^{\mu}-QF^{\mu\nu}\partial^{p}_{\nu}$. 

We next write down an expansion of $\mathscr{J}_{\chi}^{\mu}$ also in powers of $\hbar$, i.e
\begin{equation}\label{eq:075}
\mathscr{J}_{\chi,\mu}=\mathscr{J}^{(0)}_{\chi,\mu}+\hbar\mathscr{J}^{(1)}_{\chi,\mu}+\hbar^2\mathscr{J}^{(2)}_{\chi,\mu}+\mathbf{O}(\hbar^{3}).
\end{equation} 
By substituting   the operators in Eq.~(\ref{eq:074}) and chiral component in Eq.~(\ref{eq:075}) into the Eqs. (\ref{eq.CKE.1}-\ref{eq.CKE.3}) , one obtains: 
\begin{eqnarray}
0&=&\Big[p_{\mu}\mathscr{J}^{(0)}_{\chi,\nu}-p_{\nu}\mathscr{J}^{(0)}_{\chi,\mu}\Big]
+\hbar\Big[
\epsilon_{\mu\nu\rho\sigma}\triangledown^{\rho}\mathscr{J}^{(0)\sigma}_{\chi} + 2\chi(p_{\mu}\mathscr{J}^{(1)}_{\chi,\nu}-p_{\nu}\mathscr{J}^{(1)}_{\chi,\mu})
 \Big] + \mathbf{O}(\hbar^{2}),
\label{eq.CKE.expand.1}\\
0&=&\Big[p^{\mu}\mathscr{J}^{(0)}_{\chi,\mu}\Big]
+\hbar\Big[ p^{\mu}\mathscr{J}^{(1)}_{\chi,\mu} \Big] +\mathbf{O}(\hbar^{2}),
\label{eq.CKE.expand.2}\\
0&=&\Big[\triangledown^{\mu}\mathscr{J}^{(0)}_{\chi,\mu}\Big]
+\hbar\Big[ \triangledown^{\mu}\mathscr{J}^{(1)}_{\chi,\mu} \Big] +\mathbf{O}(\hbar^{2}).
\label{eq.CKE.expand.3}
\end{eqnarray}
Just as the strategy in perturbation theory, one can then match the terms in the above equations at each given order of $\hbar$ and obtain an infinite series of equations order by order. In this paper we will only deal with the two leading orders, i.e. the order $\hbar^0$ equations and the order $\hbar^1$ equations.

Let us first examine  the zeroth order equations: 
\begin{eqnarray}
0&=&p_{\mu}\mathscr{J}^{(0)}_{\chi,\nu}-p_{\nu}\mathscr{J}^{(0)}_{\chi,\mu},
\label{eq.CKE.0.1}\\
0&=&p^{\mu}\mathscr{J}^{(0)}_{\chi,\mu},
\label{eq.CKE.0.2}\\
0&=&\triangledown^{\mu}\mathscr{J}^{(0)}_{\chi,\mu}.
\label{eq.CKE.0.3}
\end{eqnarray}
%Up to the first order $\triangledown_{\mu}=\partial_{\mu}-QF_{\mu\nu}\partial_{p}^{\nu}$.
Eqs. (\ref{eq.CKE.0.1}) \& (\ref{eq.CKE.0.2}) are the constraint conditions for the current $\mathscr{J}^{(0)}_{\chi,\mu}$: the former requires that $\mathscr{J}^{(0)}_{\chi,\mu}$ must be parallel to $p_\mu$, {\it i.e.} $\mathscr{J}^{(0)}_{\chi,\mu}=p_\mu S(x,p)$ where $S$ is certain arbitrary scalar function; the latter further demands that $p^2S(x,p)=0$. These conditions uniquely fix the general form of  the zeroth order current to be the following: 
\begin{align}\begin{split}\label{eq:066}
&\mathscr{J}_{\mu,\chi}^{(0)}=p_{\mu}f^{(0)}_{\chi}\delta(p^2),  \end{split}
\end{align}
together with the classical on-shell condition as reflected in the delta-function. Apparently $f^{(0)}_{\chi}$ is the classical phase-space distribution function, which can be further decomposed as: 
\begin{align}\begin{split}\label{eq:066b}
&f^{(0)}_{\chi}(x,p)=\sum_{\epsilon=\pm 1}\theta(\epsilon p^{0})f^{(0)\epsilon}_{\chi}(x,\epsilon p).
\end{split}
\end{align}
where $\epsilon=\pm1$ corresponds to particle with positive/negative energy.  
%Comparing the chiral current in Eq. (\ref{eq:067}) with the definition of current density $j^{\mu}(x)=2\int d^{4}p~\delta(p^{2}) p^{\mu}f(x,p)$,
%one can see clearly that the $f^{(0)}_{\chi}$ term in Eq.(\ref{eq:066}) means exactly the zeroth order chiral distribution function in eight-dimensional phase space $(x,p)$.

Finally, by substituting Eq.(\ref{eq:066}) into the evolution equation (\ref{eq.CKE.0.3}), one obtains the zeroth order transport equation,
\begin{equation}\label{eq:041}
\delta(p^2) p^{\mu} \triangledown_{\mu} f^{(0)}_{\chi}=
\delta(p^2) p^{\mu}(\partial_{\mu}-QF_{\mu\nu}\partial^{\nu}_{p}) f^{(0)}_{\chi}=0,
\end{equation}
which is the classical covariant Vlasov equation.

%============================================================================
\subsection{The $\hbar$-order constraint equations and general solutions}\label{sec.hhbar}

We now  move on to examine the first order equations, as follows:
\begin{eqnarray}
0&=&\epsilon_{\mu\nu\rho\sigma}\triangledown^{\rho}\mathscr{J}^{(0)\sigma}_{\chi} + 2\chi(p_{\mu}\mathscr{J}^{(1)}_{\chi,\nu}-p_{\nu}\mathscr{J}^{(1)}_{\chi,\mu}), \label{eq.CKE.1.1}\\
0&=&p^{\mu}\mathscr{J}^{(1)}_{\chi,\mu}, \label{eq.CKE.1.2}\\
0&=&\triangledown^{\mu}\mathscr{J}^{(1)}_{\chi,\mu}.\label{eq.CKE.1.3}
\end{eqnarray}
Here, Eq.(\ref{eq.CKE.1.1}) gives the connection between the zeroth and first order of $\mathscr{J}_{\mu}$.
Noting that
\begin{eqnarray*}
\triangledown^{\rho}\mathscr{J}^{(0)\sigma}_{\chi}&=-QF^{\rho\sigma}f^{(0)}_{\chi}\delta(p^2)+p^{\sigma}\left(\triangledown^{\rho}f^{(0)}_{\chi}\right)\delta(p^2)-2QF^{\rho\lambda}p_{\lambda}p^{\sigma}f^{(0)}_{\chi}\delta^{'}(p^2),
\end{eqnarray*}
and using the Schouten identity
\begin{eqnarray}\label{eq.schouten}
&p_{\mu}\epsilon_{\nu\rho\sigma\lambda}+p_{\nu}\epsilon_{\rho\sigma\lambda\mu}+p_{\rho}\epsilon_{\sigma\lambda\mu\nu}+p_{\sigma}\epsilon_{\lambda\mu\nu\rho}+p_{\lambda}\epsilon_{\mu\nu\rho\sigma}=0,
\end{eqnarray}
we obtain
\begin{eqnarray*}
\epsilon_{\mu\nu\rho\sigma}F^{\rho\lambda}p_{\lambda}p^{\sigma}f^{(0)}_{\chi}\delta^{'}(p^2)&=&p_{\mu}\widetilde{F}_{\nu\sigma}p^{\sigma}f^{(0)}_{\chi}\delta^{'}(p^2)-p_{\nu}\widetilde{F}_{\mu\sigma}p^{\sigma}f^{(0)}_{\chi}\delta^{'}(p^2)+p^{2}\widetilde{F}_{\mu\nu}f^{(0)}_{\chi}\delta^{'}(p^2),
\\&~&\\
\epsilon_{\mu\nu\rho\sigma}\triangledown^{\rho}\mathscr{J}^{(0)\sigma}_{\chi}
&=&-2Q\widetilde{F}_{\mu\nu}f^{(0)}_{\chi}\delta(p^2)+\epsilon_{\mu\nu\rho\sigma}p^{\sigma}\left(\triangledown^{\rho}f^{(0)}_{\chi}\right)\delta(p^2)-2Q\epsilon_{\mu\nu\rho\sigma}F^{\rho\lambda}p_{\lambda}p^{\sigma}f^{(0)}_{\chi}\delta^{'}(p^2),\\
&=&\epsilon_{\mu\nu\rho\sigma}p^{\sigma}\left(\triangledown^{\rho}f^{(0)}_{\chi}\right)\delta(p^2)-2Q\left( p_{\mu}\widetilde{F}_{\nu\sigma}-p_{\nu}\widetilde{F}_{\mu\sigma} \right)p^{\sigma}f^{(0)}_{\chi}\delta^{'}(p^2).
\end{eqnarray*}
Here $\delta^{'}(p^{2})=d\delta(p^2)/dp^2$, and we have used the relation $p^{2}\delta^{'}(p^{2})=-\delta(p^{2})$.
Now Eq.(\ref{eq.CKE.1.1}) becomes: 
\begin{align}\label{eq:068}
&\epsilon_{\mu\nu\rho\sigma}p^{\sigma}\left(\triangledown^{\rho}f^{(0)}_{\chi}\right)\delta(p^2)-2Q\left( p_{\mu}\widetilde{F}_{\nu\sigma}-p_{\nu}\widetilde{F}_{\mu\sigma} \right)p^{\sigma}f^{(0)}_{\chi}\delta^{'}(p^2)=-2\chi(p_{\mu}\mathscr{J}^{(1)}_{\nu,\chi}-p_{\nu}\mathscr{J}^{(1)}_{\mu,\chi}).
\end{align}
Contracting both sides of the above equation with $p^{\nu}$ and using Eq.(\ref{eq.CKE.1.2}), one can derive that
\begin{align}
&p^{2}\left( Q\widetilde{F}_{\mu\sigma}p^{\sigma}f^{(0)}_{\chi}\delta^{'}(p^2)- \chi\mathscr{J}^{(1)}_{\mu,\chi} \right)=0.
\end{align}
hence the most general solution to the above constraint equation can be written as,
\begin{align}
&\mathscr{J}^{(1)}_{\mu,\chi}=\mathscr{H}_{\mu}\delta(p^{2})+\chi Q \widetilde{F}_{\mu\nu}p^{\nu}f^{(0)}_{\chi}\delta^{'}(p^2).
\end{align}
In the above, the $\mathscr{H}_{\mu}$ is an arbitrary Lorentz vector. By using the second constraint  Eq.(\ref{eq.CKE.1.2}), one arrives at: 
\begin{align}
&p^{\mu}\mathscr{H}_{\mu}\delta(p^{2})=0.
\end{align}
Due to the special nature of light-like momentum $p^\mu=(|\mathbf{p}|,\mathbf{p})$ (as mandated by the delta-function), there are three categories of vectors that can satisfy the above equation: one parallel to $p_\mu$ itself, the other two taking the form $(0,\mathbf{K})$ with the spatial component satisfying $\mathbf{K}\cdot\mathbf{p}=0$.  Thus one can  decompose $\mathscr{H}_{\mu}$ into components that are parallel/orthogonal to the momentum $p^{\mu}$ respectively: 
\begin{align}
\mathscr{H}_{\mu}=p_{\mu}f^{(1)}_{\chi}+\mathscr{K}_{\mu}.\label{eq.Hmu}
\end{align}
Here, $f^{(1)}_{\chi}$ has the natural interpretation as the first-order correction to $f^{(0)}_{\chi}$ by comparing the above with the zeroth order Eq.(\ref{eq:066}). 
%On the other hand, the orthogonal part $\mathscr{K}^\mu$, satisfying that $p^{\mu}\mathscr{K}_{\mu}=0$, should be nonsingular at $p^{2}=0$. 
To solve the orthogonal part $\mathscr{K}^\mu$, one can substitute  the representation of $\mathscr{J}^{(1)}_{\mu,\chi}$ into Eq.(\ref{eq:068}) and get the following constraint equation for $\mathscr{K}^\mu$
\begin{eqnarray}\label{eq:069}
\epsilon_{\mu\nu\rho\sigma}p^{\sigma}\left(\triangledown^{\rho}f^{(0)}_{\chi}\right)\delta(p^2)=-2\chi(p_{\mu}\mathscr{K}_{\nu}-p_{\nu}\mathscr{K}_{\mu})\delta(p^{2}).
\end{eqnarray}
The most general solution of $\mathscr{K}_{\mu}$ can be expressed as
\begin{eqnarray}
\mathscr{K}_{\mu}=\frac{\chi}{2p\cdot n}\epsilon_{\mu\nu\lambda\rho}p^{\nu}n^{\lambda}\left(\triangledown^{\rho}f^{(0)}_{\chi}\right),\label{eq.Kmu}
\end{eqnarray}
where  an {\it arbitrary} auxiliary time-like unit vector $n^\mu$ (satisfying $n^\mu n_\mu=1$) has been introduced. It should be noted that the above is the correct solution to the constraint equations even for spacetime dependent $n^\mu(x)$. A detailed proof of this solution is included in the Appendix \ref{sec.appendix.Kmu}. The meaning of $n^\mu$ and the pertinent frame dependence issue will be discussed in the next subsection.

Finally we can combine the solutions to the zeroth and first order constraint equations, and write down the following expression of $\mathscr{J}^{\mu}_{\chi}$ up to the first order of $\hbar$: 
\begin{eqnarray} \label{eq.Jhbar}
\mathscr{J}^{\mu}_\chi&=p^{\mu}f_{\chi}\delta(p^{2})+\hbar\chi Q\widetilde{F}^{\mu\nu}p_{\nu}f^{(0)}_{\chi}\delta^{'}(p^{2})-\hbar\frac{\chi}{2p\cdot n}\epsilon^{\mu\nu\lambda\rho}n_{\nu}p_{\lambda}\left(\triangledown_{\rho}f^{(0)}_{\chi}\right)\delta(p^2).
\end{eqnarray}
Here $\widetilde{F}^{\mu\nu}=\frac{1}{2}\epsilon_{\mu\nu\rho\sigma}F^{\rho\sigma}$ is the dual tensor of $F^{\mu\nu}$. We have introduced the distribution function $f_{\chi}$ including $\hbar$-order quantum correction: 
 $$f_{\chi}=f^{(0)}_{\chi}+\hbar f^{(1)}_{\chi}$$ 
 which can also be decomposed into positive/negative energy parts, like Eq.(\ref{eq:066}),
$f_{\chi}(x,p)=\sum_{\epsilon=\pm 1}\theta(\epsilon p^{0})f^{\epsilon}_{\chi}(x,\epsilon p)$.
Now the chiral current is given by 
\begin{eqnarray}
J^{\mu}_{\chi}&=\int d^{4}p \mathscr{J}^{\mu}_{\chi}=J^{(0)\mu}_{\chi}+\hbar J^{(1)\mu}_{\chi}.
\end{eqnarray}
with  the zeroth order $J^{(0)\mu}_{\chi}$ and first order  $J^{(1)\mu}_{\chi}$   expressed as
\begin{align}\begin{split}
&J^{(0)\mu}_{\chi}=\int d^{4}p p^{\mu}f^{(0)}_{\chi}\delta(p^{2}),\\
&J^{(1)\mu}_{\chi}=\int d^{4}p p^{\mu}f^{(1)}_{\chi}\delta(p^{2})+\chi Q\widetilde{F}^{\mu\nu}\int d^{4}p p_{\nu}f^{(0)}_{\chi}\delta^{'}(p^2)-\frac{\chi}{2}\epsilon^{\mu\nu\lambda\rho}n_{\nu} \int d^{4}p \frac{1}{p\cdot n}p_{\lambda}\left(\triangledown_{\rho}f^{(0)}_{\chi}\right)\delta(p^2) \,\, .
\end{split}\end{align}
Similarly, one can also get the expression of the vector/axial currents and energy-momentum tensor from $\mathscr{J}^{\mu}_\chi$.

%============================================================================
\subsection{Lorentz invariance and frame dependence}\label{sec.frame}

In the solution for $\mathscr{K}^\mu$ in Eq.(\ref{eq.Kmu}), an arbitrary auxiliary quantity $n^\mu$ that appears to be a free choice at our disposal without clear physical meaning. A more careful examination reveals that the quantity $n^\mu$ actually plays a subtle yet crucial role in the chiral transport, especially pertaining to the  confusing issues of Lorentz invariance and frame dependence, which we shall discuss next. 

  To understand the role of $n^\mu$, let us come back to the decomposition of $\mathscr{H}_{\mu}$ in Eq.(\ref{eq.Hmu}), i.e.  $\mathscr{H}_{\mu}= p_{\mu}f^{(1)}_{\chi} + \mathscr{K}_{\mu}$. As already mentioned above and as discussed with great details in the Appendix \ref{sec.appendix.Kmu}, this decomposition is subtle due to the light-like nature of the $p^\mu$.  To {\em unambiguously} identify the first order correction to the distribution function, one must demand that  the part along $p_\mu$ should be attributed to the distribution term $f^{(1)}_{\chi}$ while the rest to the $\mathscr{K}_{\mu}$ term. In fact, such a requirement completely fixes the form of $\mathscr{K}_{\mu}$. 
  %For the light-like vector $p_\mu=(|\mathbf{p}|,\mathbf{p})$, there are three categories of orthogonal vectors: one parallel to $p_\mu$ itself, the other two taking the form $(0,\mathbf{K})$ with the spatial component satisfying $\mathbf{K}\cdot\mathbf{p}=0$. Therefore, 
  For a uniquely defined $f^{(1)}_{\chi}$, the $\mathscr{K}_{\mu}$ must take the form $(0,\mathbf{K})$ with the spatial component satisfying $\mathbf{K}\cdot\mathbf{p}=0$.  Combining this requirement with Eq.(\ref{eq.Kmu}), one arrives at the unique choice $n^\mu=(1,0,0,0)$ and the corresponding   $\mathscr{K}_{\mu}$  below: 
 \begin{eqnarray} \label{eq:Krest}
\mathscr{K}_{\mu} = \left ( 0,  \frac{\chi}{2|\mathbf{p}|}  \mathbf{p} \times 
(\vec{\triangledown}f^{(0)}_{\chi})   \right ) \,\, .
 \end{eqnarray}

This however is not the end of the story. While the above construction gives  well-defined $f^{(1)}_{\chi} p_\mu$ and $\mathscr{K}_{\mu}$  in the {\em current reference frame, this decomposition is actually  frame dependent}. To appreciate this less obvious subtlety,  suppose in the current frame there is a vector $\mathscr{K}_\mu=(0,\mathbf{K})$ which satisfies orthogonality to $p_\mu$ via $\mathbf{K}\cdot \mathbf{p}=0$. But upon boosting into a different frame with both $p_\mu$ and $\mathscr{K}_\mu$ transformed as Lorentz vectors into 
$p'$ and $\mathscr{K}'$, one finds that in general $\mathscr{K}'$ acquires a component along $p'$, despite that they still satisfy $\mathscr{K}' \cdot p'=0$. That means one has to redo the proper decomposition in the new reference frame and find a different $\mathscr{K}''=(0,\mathbf{K}'')$ satisfying $\mathbf{K}''\cdot p'=0$. This issue again arises from the light-like nature of $p_\mu$.    
 
A lengthy calculation in the Appendix \ref{sec.appendix.Kmu} proves that if one boosts from the current frame to a different frame of four-velocity $u^\mu$ (with respect to the current frame), then the  $\mathscr{K}_{\mu}$ from proper decomposition in this new frame should be precisely and uniquely given by Eq.(\ref{eq.Kmu}) with the identification $n^\mu \to u^\mu$ which leaves a well-defined $f^{(1)}_{\chi}$ in this new reference frame. Hence the role of $n^\mu$ now becomes clear. This result also explicates the fact that the distribution term $f^{(1)}_{\chi}$ becomes {\em frame-dependent} as well.  While the distribution function  in usual transport theory is a Lorentz scalar, 
here it is demonstrated clearly that in chiral transport theory a nontrivial frame dependence of the distribution function arises precisely at the $\hat{\mathbf{O}}(\hbar)$ order correction and in the specific way discussed above.

In short, the Wigner function formalism is in itself totally covariant and  it is the decomposition of $\mathscr{H}_{\mu}$ that introduces frame dependence. The unique identification of $f^{(1)}_{\chi} $ requires the $\mathscr{K}^\mu$ to contain no $p^\mu$-parallel component while this requirement is frame-dependent. For an observer with velocity $u^\mu=n^\mu$, the Eq.(\ref{eq.Kmu}) gives the correct $\mathscr{K}^\mu$. The peculiar structure of $\mathscr{K}^\mu$  also clarifies  the frame dependence of spin tensor $S^{\mu\nu}$ and the side-jump effect ~\cite{Chen:2015gta,Hidaka:2016yjf,Hidaka:2017auj}.

%============================================================================
\section{The Covariant Chiral Transport Equation}\label{sec:3}

%============================================================================
\subsection{Covariant chiral transport equation}

In this subsection, we focus on deriving the covariant chiral transport equation up to $\hbar$ order, which can be obtained by substituting Eq.(\ref{eq.Jhbar}) into   Eq.(\ref{eq.CKE.3}): 
\begin{eqnarray}
0&=&\triangledown_{\mu}\mathscr{J}^{\mu}_\chi\nonumber\\
&=&\triangledown_{\mu}\left(p^{\mu}f_{\chi}\delta(p^{2})\right)+\hbar\chi Q\triangledown_{\mu}\left(\widetilde{F}^{\mu\nu}f^{(0)}_{\chi}p_{\nu}\delta^{'}(p^{2})\right)-\hbar\frac{\chi}{2}\epsilon^{\mu\nu\lambda\rho}\triangledown_{\mu}\left(\frac{1}{p\cdot n}n_{\nu}p_{\lambda}\left(\triangledown_{\rho}f^{(0)}_{\chi}\right)\delta(p^2)\right).
\end{eqnarray}
One can further simplify the first term of the above equation as
\begin{eqnarray*}
\triangledown_{\mu}\left(p^{\mu}f_{\chi}\delta(p^{2})\right)=\delta(p^{2})p\cdot\triangledown f_{\chi},
\end{eqnarray*}
and the second term as
\begin{eqnarray*}
&&\triangledown_{\mu}\left(\widetilde{F}^{\mu\nu}f^{(0)}_{\chi}p_{\nu}\delta^{'}(p^{2})\right)\\
&=&-Q\widetilde{F}^{\mu\nu}F_{\mu\nu}f^{(0)}_{\chi}\delta^{'}(p^{2})-2Q\widetilde{F}^{\mu\nu}p_{\nu}F_{\mu\lambda}p^{\lambda}f^{(0)}_{\chi}\delta^{''}(p^{2})+\widetilde{F}^{\mu\nu}p_{\nu}\left(\triangledown_{\mu}f^{(0)}_{\chi}\right)\delta^{'}(p^{2})\\
&=&-Q\widetilde{F}^{\mu\nu}F_{\mu\nu}f^{(0)}_{\chi}\delta^{'}(p^{2})-\frac{1}{2}Q\widetilde{F}^{\mu\nu}F_{\mu\nu}p^{2}f^{(0)}_{\chi}\delta^{''}(p^{2})+\widetilde{F}^{\mu\nu}p_{\nu}\left(\triangledown_{\mu}f^{(0)}_{\chi}\right)\delta^{'}(p^{2})\\
&=&\widetilde{F}^{\mu\nu}p_{\nu}\left(\triangledown_{\mu}f^{(0)}_{\chi}\right)\delta^{'}(p^2),
\end{eqnarray*}
while the third term as
\begin{eqnarray*}
&&\epsilon^{\mu\nu\lambda\rho}\triangledown_{\mu}\left(\frac{1}{p\cdot n}n_{\nu}p_{\lambda}\left(\triangledown_{\rho}f^{(0)}_{\chi}\right)\delta(p^2)\right)\\
&=&2Q\widetilde{F}^{\rho\lambda}p_{\lambda}\left(\triangledown_{\rho}f^{(0)}_{\chi}\right)\delta^{'}(p^2)+\frac{2Q}{p\cdot n}p_{\lambda}\widetilde{F}^{\lambda\nu}n_{\nu}\delta^{'}(p^2)p\cdot\triangledown f^{(0)}_{\chi}\\
&&-\frac{1}{\left(p\cdot n\right)^2}\left[(\partial_{\mu}n_{\sigma})p^{\sigma}-QF_{\mu\alpha}n^{\alpha}\right]\epsilon^{\mu\nu\lambda\rho}n_{\nu}p_{\lambda}\left(\triangledown_{\rho}f^{(0)}_{\chi}\right)\delta(p^2)\\
&&+\frac{1}{p\cdot n}\epsilon^{\mu\nu\lambda\rho}\left(\partial_{\mu}n_{\nu}\right) p_{\lambda}\left(\triangledown_{\rho}f^{(0)}_{\chi}\right)\delta(p^2)-\frac{Q}{p\cdot n}p_{\lambda}\left(\partial_{\sigma}\widetilde{F}^{\lambda\nu}\right)n_{\nu}\left(\partial^{\sigma}_{p} f^{(0)}_{\chi}\right)\delta(p^2).
\end{eqnarray*}
In the above steps, we have used the relation $p^{2}\delta^{'}(p^2)=-\delta(p^2)$, $p^{2}\delta^{''}(p^2)=-2\delta^{'}(p^2)$,
the Bianchi identity $\triangledown_{\mu}\widetilde{F}^{\mu\nu}=\partial_{\mu}\widetilde{F}^{\mu\nu}=0$,
and $4\widetilde{F}^{\mu\nu}p_{\nu}F_{\mu\alpha}p^{\alpha}=p^{2}\widetilde{F}^{\mu\nu}F_{\mu\nu}$,
which can be easily proved by the Schouten identity Eq.(\ref{eq.schouten}).
Also, we have used the following relations,
\begin{eqnarray*}
&&\epsilon^{\mu\nu\lambda\rho}\left(\triangledown_{\mu}\triangledown_{\rho}f^{(0)}_{\chi}\right)=\frac{1}{2}\epsilon^{\mu\nu\lambda\rho}[\triangledown_{\mu},\triangledown_{\rho}]f^{(0)}_{\chi}=Q\left(\partial_{\sigma}\widetilde{F}^{\nu\lambda}\right)\partial^{\sigma}_{p} f^{(0)}_{\chi},\\
&&2p^{\alpha}\epsilon^{\mu\nu\lambda\rho}n_{\nu}p_{\lambda}F_{\mu\alpha}=-2(p\cdot n)\widetilde{F}^{\rho\lambda}p_{\lambda}+2p^{2}\widetilde{F}^{\rho\nu}n_{\nu}-2p_{\lambda}\widetilde{F}^{\lambda\nu}n_{\nu}p_{\rho}.
\end{eqnarray*}

Finally, we obtain the following covariant Chiral Kinetic Equation as the evolution equation for the  distribution function $f_\chi$ up to $\hbar$-order quantum correction: 
\begin{eqnarray}\label{eq.cCKE}
0&=&\triangledown_{\mu}\mathscr{J}^{\mu}\nonumber\\
&=&\delta(p^{2})p\cdot\triangledown f_{\chi}-\hbar\frac{\chi Q}{p\cdot n}p_{\lambda}\widetilde{F}^{\lambda\nu}n_{\nu}\delta^{'}(p^{2}) p\cdot\triangledown f^{(0)}_{\chi}\nonumber\\
&&+\hbar\frac{\chi}{2\left(p\cdot n\right)^2}\left[(\partial_{\mu}n_{\sigma})p^{\sigma}-QF_{\mu\alpha}n^{\alpha}\right]\epsilon^{\mu\nu\lambda\rho}n_{\nu}p_{\lambda}\left(\triangledown_{\rho}f^{(0)}_{\chi}\right)\delta(p^2)\nonumber\\
&&-\hbar\frac{\chi}{2p\cdot n}\epsilon^{\mu\nu\lambda\rho}\left(\partial_{\mu}n_{\nu}\right) p_{\lambda}\left(\triangledown_{\rho}f^{(0)}_{\chi}\right)\delta(p^2)+\hbar\frac{\chi Q}{2p\cdot n}p_{\lambda}\left(\partial_{\sigma}\widetilde{F}^{\lambda\nu}\right)n_{\nu}\left(\partial^{\sigma}_{p} f^{(0)}_{\chi}\right)\delta(p^2)\nonumber\\
&=&\delta\left( p^{2}-\hbar\frac{\chi Q}{p\cdot n}p_{\lambda}\widetilde{F}^{\lambda\nu}n_{\nu} \right) \Bigg\{ p\cdot\triangledown +\hbar\frac{\chi}{2\left(p\cdot n\right)^2}\left[(\partial_{\mu}n_{\sigma})p^{\sigma}-QF_{\mu\alpha}n^{\alpha}\right]\epsilon^{\mu\nu\lambda\rho}n_{\nu}p_{\lambda}\triangledown_{\rho} \nonumber\\
&&\qquad\qquad\qquad\qquad\qquad\qquad -\hbar\frac{\chi}{2p\cdot n}\epsilon^{\mu\nu\lambda\rho}\left(\partial_{\mu}n_{\nu}\right) p_{\lambda}\triangledown_{\rho}+\hbar\frac{\chi Q}{2p\cdot n}p_{\lambda}\left(\partial_{\sigma}\widetilde{F}^{\lambda\nu}\right)n_{\nu}\partial^{\sigma}_{p} \Bigg\} f_{\chi}.
\end{eqnarray}
In the last step we have  used the Taylor expansion in $\delta$ function and we only keep terms up to the $\hbar$ order. 
One can see from the argument of the delta function that the energy of chiral particle has been shifted in $\hbar$ order, showing the effect of quantum correction.
Eq.(\ref{eq.cCKE}) is the complete and consistent covariant chiral kinetic equation. Notably, the mass shell condition in the delta-function has shifted from the classical case and receives an $\hbar$-order quantum correction which has the physical interpretation of magnetization energy due to interaction between the charged chiral fermion's magnetic moment with the external magnetic field. Again, it's worth emphasizing that the expression of distribution function $f_{\chi}$, or more strictly speaking the first order correction $f^{(1)}_{\chi}$, depends on the choice of $n^{\mu}$.

%============================================================================
\subsection{3D Chiral Kinetic Equation }\label{sec.3dCKT}

In this subsection, let's consider a simplified case and take $n^{\mu}$ as a constant-homogeneous 4-vector $u^{\mu}$.
In this case, the Eq.(\ref{eq.cCKE})  can be written as
\begin{eqnarray}\label{eq:083}
0&=&\delta\left(p^{2}-\hbar\frac{\chi Q}{p\cdot u}(B\cdot p)\right)\left\{p^{\rho}\triangledown_{\rho} -\hbar\frac{\chi Q}{2\left(p\cdot u\right)^2}\epsilon^{\mu\nu\lambda\rho}E_{\mu} u_{\nu}p_{\lambda}\triangledown_{\rho}+\hbar\frac{\chi Q}{2p\cdot u}p_{\lambda}\left(\partial_{\rho}B^{\lambda}\right)\partial^{\rho}_{p} \right\}f_{\chi},
\end{eqnarray}
where we introduce the notations $E^{\mu}=F^{\mu\nu}u_{\nu}, \quad B^{\mu}=\widetilde{F}^{\mu\nu}u_{\nu}$. In addition, using the following relations, 
\begin{eqnarray*}
&&F^{\mu\nu}=E^{\mu}u^{\nu}-E^{\nu}u^{\mu}+\epsilon^{\mu\nu\rho\sigma}u_{\rho}B_{\sigma},\nonumber\\
&&\epsilon^{\mu\nu\lambda\rho}\epsilon_{\rho\alpha\beta\delta}=\delta^{\mu}_{\alpha}\delta^{\nu}_{\beta}\delta^{\lambda}_{\delta}+\delta^{\mu}_{\beta}\delta^{\nu}_{\delta}\delta^{\lambda}_{\alpha}+\delta^{\mu}_{\delta}\delta^{\nu}_{\alpha}\delta^{\lambda}_{\beta}-\delta^{\nu}_{\alpha}\delta^{\mu}_{\beta}\delta^{\lambda}_{\delta}-\delta^{\nu}_{\beta}\delta^{\mu}_{\delta}\delta^{\lambda}_{\alpha}-\delta^{\nu}_{\delta}\delta^{\mu}_{\alpha}\delta^{\lambda}_{\beta},
\end{eqnarray*}
Eq.(\ref{eq:083}) can be reduced to
\begin{align}\label{eq:058}
0&=\delta\left(p^{2}-\hbar\frac{\chi Q}{p\cdot u}(B\cdot p)\right)\left\{\left( p^{\rho}-\hbar\frac{\chi Q}{2\left(p\cdot u\right)^2}\epsilon^{\mu\nu\lambda\rho}E_{\mu} u_{\nu}p_{\lambda}\right)\partial_{\rho} \right.\nonumber\\
&\left.~~~~~~~~~~~~~~~~~~~~~~~~~~~~~~~~ +Q\Bigg[ -\left(E\cdot p\right)u^{\rho}+(p\cdot u)E^{\rho}+\epsilon^{\mu\nu\rho\sigma}p_{\mu}u_{\nu}B_{\sigma}\right.\nonumber\\
&\left.~~~~~~~~~~~~~~~~~~~~~~~~~~~~~~~~ +\hbar\frac{\chi Q}{2\left(p\cdot u\right)^2}\left( (B\cdot p)E^{\rho}-(E\cdot B)\bar{p}^{\rho}\right) +\hbar\frac{\chi }{2p\cdot u}p_{\lambda}\left(\partial^{\rho}B^{\lambda}\right) \Bigg]\partial_{\rho}^{p}  \right\}f_{\chi}.
\end{align}

One could further simplify the above equation by choosing $n^\mu=u^\mu=(1,0,0,0)$, which can be achieved by a proper Lorentz transformation.
In this frame, $E^{\mu}=(0,\mathbf{E}), B^{\mu}=(0,\mathbf{B})$, and $\bar{p}^{\rho}=(0,\bf{p})$ is the three momentum, $p\cdot u=p_{0}$ is the energy, while $B\cdot p=-\mathbf{B}\cdot \mathbf{p}$, $E\cdot p=-\mathbf{E}\cdot \mathbf{p}$, $E\cdot B=-\mathbf{E}\cdot \mathbf{B}$.
%Eq.(\ref{eq:058}) is the 4-d chiral kinetic equation in the rest frame.
From the delta function of Eq.(\ref{eq:058}), we can get the shifted energy in external field up to $\hbar$-order: 
\begin{align}
p_{0}=\epsilon|\mathbf{p}|\left( 1-\hbar\epsilon Q  \mathbf{B}\cdot \bf{b}_{\chi}\right)=\epsilon E_{\mathbf{p}},
\quad E_{\mathbf{p}}=|\mathbf{p}|\left( 1-\hbar\epsilon Q  \mathbf{B}\cdot \bf{b}_{\chi}\right),\label{eq.onshell}
\end{align}
where $\mathbf{b}_{\chi}=\chi\frac{\mathbf{p}}{2|\mathbf{p}|^{3}}$ is the Berry curvature, $\widehat{\mathbf{p}}=\mathbf{p}/|\mathbf{p}|$ is the unit vector of momentum and $\epsilon=\pm1$ correspond to the particle with positive/negative energy.
With the shifted energy, the group velocity of the quasi-particle becomes
\begin{align}
\widetilde{\mathbf{v}}=\frac{\partial E_{\mathbf{p}}}{\partial\mathbf{p}}=\widehat{\mathbf{p}}\left(1+2\hbar\epsilon Q\mathbf{B}\cdot\mathbf{b}_{\chi} \right)-\hbar\epsilon Q b_{\chi}\mathbf{B}.
\end{align}

Note that the on-shell condition Eq.(\ref{eq.onshell}) constrains the energy  in Eq.(\ref{eq:058}) hence it's no longer a free variable in the distribution function. By integrating  Eq. (\ref{eq:058}) over $p_0$, one arrives at the following 3-dimensional chiral kinetic equation: 
\begin{align}\begin{split}
&\sum_{\epsilon=\pm 1}\frac{\epsilon E_{\mathbf{p}}}{2}\Bigg\{ \frac{1}{E_{\mathbf{p}}}\partial_{t}+\epsilon\left(\frac{p^{k}}{E^{2}_{\mathbf{p}}}+\hbar\frac{\chi Q}{2E_{\mathbf{p}}^{4}}\epsilon^{ijk}E_{i}p_{j} \right)\partial_{k}\\
&+Q\left[ \frac{E^{k}}{E_{\mathbf{p}}}+\epsilon\epsilon^{ijk}\frac{p_{i}}{E^{2}_{\mathbf{p}}}B_{j}-\hbar\epsilon\frac{\chi Q}{2E_{\mathbf{p}}^{4}}\left((\mathbf{B}\cdot \mathbf{p})E^{k}-(\mathbf{E}\cdot\mathbf{B})p^{k} \right) -\hbar\frac{\chi}{2E_{\mathbf{p}}^{3}}(\partial^{k}\mathbf{B}\cdot\mathbf{p})\right]\partial^{p}_{k}  \\
&+Q\left[-E^{k}\frac{p_{k}}{E^{2}_{\mathbf{p}}}-\hbar\epsilon\frac{\chi}{2E_{\mathbf{p}}^{3}}(\partial_{t}\mathbf{B}\cdot\mathbf{p})\right]\partial_{E_{\mathbf{p}}} \Bigg\}f^{\epsilon}_{\chi}(x,E_{\mathbf{p}},\epsilon\mathbf{p})=0.
\end{split}\end{align}
By expanding various powers of the energy $E_{\mathbf{p}}$ in $\hbar$ and keeping terms up to the first order, one obtains 
\begin{align}\begin{split}\label{eq:076}
&\sum_{\epsilon=\pm 1}\frac{\epsilon }{2}\left( 1-\hbar\epsilon Q  \mathbf{B}\cdot \bf{b}_{\chi}\right)\Bigg\{ \left( 1+\hbar\epsilon Q\mathbf{B}\cdot\mathbf{b}_{\chi}\right)\partial_{t}+\epsilon\left(\hat{p}^{k}\left( 1+2\hbar\epsilon Q\mathbf{B}\cdot\mathbf{b}_{\chi}\right)+\hbar Q\epsilon^{ijk}E_{i}b_{\chi j} \right)\partial_{k}\\
&+Q\left[E^{k}+\frac{1}{\epsilon Q}(\partial^{k}E_{\mathbf{p}})+\epsilon\epsilon^{ijk}\hat{p}_{i}B_{j}\left( 1+2\hbar\epsilon Q\mathbf{B}\cdot\mathbf{b}_{\chi}\right)+\hbar\epsilon Q\left(\mathbf{E}\cdot\mathbf{B}\right)b^{k}_{\chi}\right]\partial^{p}_{k}  \\
&+\left[ -Q E^{k}\cdot\hat{p}_{k}\left( 1+2\hbar\epsilon Q\mathbf{B}\cdot\mathbf{b}_{\chi}\right)+(\partial_{t}E_{\mathbf{p}})\Bigg]\partial_{E_{\mathbf{p}}} \right\}f^{\epsilon}_{\chi}(x,E_{\mathbf{p}},\epsilon\mathbf{p})=0.
\end{split}\end{align} 

The next step is to turn  the energy-derivative terms into the derivative terms with respect to the actual independent variables (i.e. spacetime coordinates and three-momentum):
\begin{align}\begin{split}\label{eq:077}
&\sum_{\epsilon=\pm 1}\frac{\epsilon }{2}\left( 1-\hbar\epsilon Q  \mathbf{B}\cdot \bf{b}_{\chi}\right)\Bigg\{ \left( 1+\hbar\epsilon Q\mathbf{B}\cdot\mathbf{b}_{\chi}\right)\left(\partial_{t}+(\partial_{t}E_{\mathbf{p}})\partial_{E_{\mathbf{p}}}\right)\\
&+\epsilon\Bigg(\hat{p}^{k}\left( 1+2\hbar\epsilon Q\mathbf{B}\cdot\mathbf{b}_{\chi}\right)+\hbar Q\epsilon^{ijk}E_{i}b_{\chi j} \Bigg)\left(\partial_{k}+(\partial_{k}E_{\mathbf{p}})\partial_{E_{\mathbf{p}}}\right)\\
&+Q\Bigg[E^{k}+\frac{1}{\epsilon Q}(\partial^{k}E_{\mathbf{p}})+\epsilon\epsilon^{ijk}\hat{p}_{i}B_{j}\left( 1+2\hbar\epsilon Q\mathbf{B}\cdot\mathbf{b}_{\chi}\right)-\hbar\epsilon Q\left(\mathbf{E}\cdot\mathbf{B}\right)b^{k}_{\chi}\Bigg]\left(\partial^{p}_{k}+(\partial^{p}_{k}E_{\mathbf{p}})\partial_{E_{\mathbf{p}}}\right)  \Bigg\}f^{\epsilon}_{\chi}(x,E_{\mathbf{p}},\epsilon\mathbf{p})=0.
\end{split}\end{align}
Using the expression that $\widetilde{v}_{k}=-\partial^{p}_{k}E_{\mathbf{p}}$, Eq.(\ref{eq:077}) can be further simplified,
\begin{align}\begin{split}\label{eq:078}
&\sum_{\epsilon=\pm 1}\frac{\epsilon }{2}\left( 1-\hbar\epsilon Q  \mathbf{B}\cdot \bf{b}_{\chi}\right)\Bigg\{ \left( 1+\hbar\epsilon Q\mathbf{B}\cdot\mathbf{b}_{\chi}\right)\left(\partial_{t}+(\partial_{t}E_{\mathbf{p}})\partial_{E_{\mathbf{p}}}\right)\\
&+\epsilon\Bigg(\widetilde{v}^{k}+\hbar\epsilon Q(\widetilde{\mathbf{v}}\cdot\mathbf{b}_{\chi})B^{k} +\hbar Q\epsilon^{ijk}E_{i}b_{\chi j} \Bigg)\left(\partial_{k}+(\partial_{k}E_{\mathbf{p}})\partial_{E_{\mathbf{p}}}\right)\\
&+Q\left[E^{k}+\frac{1}{\epsilon Q}(\partial^{k}E_{\mathbf{p}})+\epsilon\epsilon^{ijk}\widetilde{v}_{i}B_{j}+\hbar\epsilon Q(\mathbf{E}\cdot\mathbf{B})b^{k}_{\chi}\right]\left(\partial^{p}_{k}+(\partial^{p}_{k}E_{\mathbf{p}})\partial_{E_{\mathbf{p}}}\right)  \Bigg\}f^{\epsilon}_{\chi}(x,E_{\mathbf{p}},\epsilon\mathbf{p})=0.
\end{split}\end{align}
%Where $\widetilde{v}_{k}=-\partial^{p}_{k}E_{\mathbf{p}}$.
%Because of that the energy $E_{\mathbf{p}}$ now is no longer an independent variable, we should use the chain rule to $(\partial_x,\partial_\mathbf{p})$, i.e
It can be seen explicitly that by employing the chain rule
\begin{eqnarray*}
{[}\partial_{t}+(\partial_{t}E_{\mathbf{p}})\partial_{E_{\mathbf{p}}}{]} f^{\epsilon}_{\chi}(t, \mathbf{x},E_{\mathbf{p}},\epsilon\mathbf{p}) &=& \partial_{t} f^{\epsilon}_{\chi}(t, \mathbf{x}, \epsilon\mathbf{p}),\\
{[}\partial_{k}+(\partial_{k}E_{\mathbf{p}})\partial_{E_{\mathbf{p}}}{]} f^{\epsilon}_{\chi}(t, \mathbf{x},E_{\mathbf{p}},\epsilon\mathbf{p}) &=& \partial_{k} f^{\epsilon}_{\chi}(t, \mathbf{x}, \epsilon\mathbf{p}),\\
{[}\partial^{p}_{k}+(\partial^{p}_{k}E_{\mathbf{p}})\partial_{E_{\mathbf{p}}}{]} f^{\epsilon}_{\chi}(t, \mathbf{x},E_{\mathbf{p}},\epsilon\mathbf{p}) &=& \partial^{p}_{k}f^{\epsilon}_{\chi}(t, \mathbf{x}, \epsilon\mathbf{p}).
\end{eqnarray*}
%Mover, let $f^{\epsilon}_{\chi}(x,E_{\mathbf{p}}, \epsilon\mathbf{p})=\frac{2}{(2\pi)^{3}}f^{\epsilon}_{\chi}(t, \mathbf{x}, \epsilon\mathbf{p})$, Eq.(\ref{eq:078}) can be reduced as
one can eventually remove the energy derivative terms and obtain: 
\begin{align}\begin{split}\label{eq:079}
&\sum_{\epsilon=\pm 1}\epsilon\Bigg\{ \partial_{t}+\epsilon\left( 1-\hbar\epsilon Q  \mathbf{B}\cdot \bf{b}_{\chi}\right)\Bigg(\widetilde{v}^{k}+\hbar\epsilon Q(\widetilde{\mathbf{v}}\cdot\mathbf{b}_{\chi})B^{k} +\hbar Q\epsilon^{ijk}E_{i}b_{\chi j} \Bigg)\partial_{k}\\
&+Q\left( 1-\hbar\epsilon Q  \mathbf{B}\cdot \bf{b}_{\chi}\right)\left[E^{k}+\frac{1}{\epsilon Q}(\partial^{k}E_{\mathbf{p}})+\epsilon\epsilon^{ijk}\widetilde{v}_{i}B_{j}+\hbar\epsilon Q(\mathbf{E}\cdot\mathbf{B})b^{k}_{\chi}\right]\partial^{p}_{k}  \Bigg\}f^{\epsilon}_{\chi}(t, \mathbf{x}, \epsilon\mathbf{p})=0,
\end{split}\end{align}
Contracting over all of index $i,j,k=1,2,3$, and replacing $\mathbf{p}$ by $\epsilon\mathbf{p}$ to convert distribution of particle with negative energy into that of anti-particle, we can write the chiral kinetic equation for particle and anti-particle separately,
\begin{align}\label{eq:065}
&\Bigg\{ \partial_{t}+\frac{1}{\sqrt{G}}\left(\widetilde{\boldsymbol{v}}+\hbar Q(\widetilde{\boldsymbol{v}}\cdot\mathbf{b}_{\chi})\mathbf{B}+\hbar Q\widetilde{\mathbf{E}}\times\mathbf{b}_{\chi}\right)\cdot\triangledown_{\mathbf{x}}+ \frac{\epsilon Q}{\sqrt{G}}\left(\widetilde{\mathbf{E}}+\widetilde{\boldsymbol{v}}\times\mathbf{B}+\hbar Q(\widetilde{\mathbf{E}}\cdot\mathbf{B})\mathbf{b}_{\chi} \right)\cdot\triangledown_{\mathbf{p}}\Bigg\}f^{\epsilon}_{\chi}(x,\mathbf{p})=0.
\end{align}
Here $\sqrt{G}=\left(1+\hbar Q\mathbf{b}_{\chi}\cdot\mathbf{B} \right)$ corresponds to the Jacobian, and
\begin{align*}
&\widetilde{\mathbf{E}}=\mathbf{E}-\frac{1}{\epsilon Q}\triangledown_{\mathbf{x}}E_{\mathbf{p}},~~~E_{\mathbf{p}}=|\mathbf{p}|\left( 1-\hbar Q  \mathbf{B}\cdot \bf{b}_{\chi}\right),~~~\widetilde{\boldsymbol{v}}=\frac{\partial E_{\mathbf{p}}}{\partial\mathbf{p}}=\widehat{\mathbf{p}}\left(1+2\hbar Q\mathbf{B}\cdot\mathbf{b}_{\chi} \right)-\hbar Q b_{\chi}\mathbf{B},
\end{align*}
where the $\chi$ denotes the chiral nor the helicity and  $f^{\epsilon}_{\chi}$ indicates the distribution function of a given chiral particle or anti-particle.
One can also convert Eq.(\ref{eq:065}) into the equation for particles with particular helicity $h\equiv\epsilon\chi$.
\begin{align}\label{eq:085}
&\Bigg\{ \partial_{t}+\frac{1}{\sqrt{G}}\left(\widetilde{\boldsymbol{v}}+\hbar\epsilon Q(\widetilde{\boldsymbol{v}}\cdot\mathbf{b}_{h})\mathbf{B}+\hbar\epsilon Q\widetilde{\mathbf{E}}\times\mathbf{b}_{h}\right)\cdot\triangledown_{\mathbf{x}}+ \frac{\epsilon Q}{\sqrt{G}}\left(\widetilde{\mathbf{E}}+\widetilde{\boldsymbol{v}}\times\mathbf{B}+\hbar\epsilon Q(\widetilde{\mathbf{E}}\cdot\mathbf{B})\mathbf{b}_{h} \right)\cdot\triangledown_{\mathbf{p}}\Bigg\}f^{\epsilon}_{h}(x,\mathbf{p})=0.
\end{align}
This reproduces   the well-known 3-dimensional chiral kinetic equation  \cite{Chen:2014cla,Hidaka:2016yjf,Hidaka:2017auj},
with the corresponding Jacobian, energy, group velocity given by 
\begin{align*}
&\sqrt{G}=\left(1+\hbar\epsilon Q\mathbf{b}_{h}\cdot\mathbf{B} \right),\quad
\widetilde{\mathbf{E}}=\mathbf{E}-\frac{1}{\epsilon Q}\triangledown_{\mathbf{x}}E_{\mathbf{p}},\\
&E_{\mathbf{p}}=|\mathbf{p}|\left( 1-\hbar\epsilon Q  \mathbf{B}\cdot \mathbf{b}_{h}\right),\quad
\widetilde{\boldsymbol{v}}=\frac{\partial E_{\mathbf{p}}}{\partial\mathbf{p}}=\widehat{\mathbf{p}}\left(1+2\hbar\epsilon Q\mathbf{B}\cdot\mathbf{b}_{h} \right)-\hbar\epsilon Q b_{h}\mathbf{B}.
\end{align*} 
Therefore  the chiral kinetic equation (\ref{eq:085}) is derived from a complete and consistent analysis of the Wigner function formalism with the semi-classical expansion method.

%=============================================================================
\section{Conclusion}
\label{sec:4}

In this paper, we've derived a covariant and complete solution  Eq.(\ref{eq.Jhbar}) for the chiral component of Wigner function, along with the corresponding   chiral transport equation (\ref{eq.cCKE}) for massless Dirac fermions, by starting from the general Wigner function formalism and carrying out a  consistent semiclassical expansion up to $\hat{\mathbf{O}}(\hbar)$ order. A detailed proof is given for the general and unique solution of the peculiar component $\mathscr{K}_{\mu}$ in the $\hat{\mathbf{O}}(\hbar)$-order chiral component of the Wigner function.  In particular, this new analysis  clarifies exactly  why and how the Lorentz invariance and frame dependence issues associated with the $\hat{\mathbf{O}}(\hbar)$ correction to the phase space distribution function arise  within a totally covariant framework. From the so-obtained chiral transport equation one also naturally derives as its consequences the 3D formulation of chiral kinetic theory as well as many special features of chiral fermions such as the magnetization energy shift, the Berry curvature, chiral anomaly, CME, etc. The covariant chiral transport theory lays a firm conceptual  foundation for describing anomalous transport in the generally non-equilibrium systems of chiral fermions.  

We end by discussing a number of  extensions and applications within the current framework. First of all, it is of great interest to explore higher order quantum effects beyond just the  $\hat{\mathbf{O}}(\hbar)$ order and in this regard the Wigner function formalism has its unique advantage. Second, it is also highly interesting to develop the equal-time quantum transport theory~\cite{Zhuang:1998kv} for chiral fermions in this framework. The 3D chiral kinetic theory only preserves the zeroth moment information of the 4D theory, and there is a whole hierarchy of equations for higher moments of the 4D theory that together forms the equal-time transport theory which turns 4D theory into a complete initial problem and is crucial for phenomenological applications. Furthermore, while we focus on the vector and axial components of the Wigner function in this paper, the other components also bear nontrivial physical meanings for physically relevant quantities such as spin density and helicity density, which could be readily studied with the same approach as here~\cite{Huang2018}. Additionally, in the current formalism it is relatively straightforward to incorporate fermion collision terms by starting from a Dirac Lagrangian including interaction terms~\cite{Zhuang:1995pd,Zhuang:1995jb}, which is also important for phenomenology. Last but not least, the role of a small nonzero mass (and generally the quantum transport of massive fermions) could be easily explored in the Wigner function formalism along similar line to the present study. These problems will be investigated in the future.

%=============================================================================
\section*{Acknowledgments}
The authors thank Jianhua Gao, Xingyu Guo, Xu-Guang Huang, Shi Pu and Qun Wang  for helpful discussions. The research of AH and PZ is supported by the NSFC and MOST Grant Nos. 11335005, 11575093, 2013CB922000 and 2014CB845400.  JL acknowledges support by the NSFC Grant No. 11735007. JL and SS are supported in part by the National Science Foundation under Grant No. PHY-1352368 and by the U.S. Department of Energy, Office of Science, Office of Nuclear Physics, within the framework of the Beam Energy Scan Theory (BEST) Topical Collaboration. YJ is supported by the startup funding of Beihang University.

\begin{appendix}
\section{Discussions on $\mathscr{K}^\mu$}\label{sec.appendix.Kmu}
In this appendix we discuss the solution of $\mathscr{K}^\mu$ (Eq.\ref{eq.Kmu}) in Sec.\ref{sec.hexpansion}.
To obtain the first order correction of the chiral vector, we need to solve the vector $\mathscr{K}^\mu$ satisfying Eq.(\ref{eq:069})
\begin{eqnarray}
\epsilon_{\mu\nu\rho\sigma} p^{\sigma}\Big(\nabla^\rho f_\chi^{(0)}\Big)\delta(p^2)
=-2\chi (p_\mu \mathscr{K}_\nu - p_\nu \mathscr{K}_\mu)\delta(p^2).
\end{eqnarray}
Let's denote $A^\rho\equiv\frac{1}{2}\chi\Big(\nabla^\rho f_\chi^{(0)}\Big)\delta(p^2)$,
and $K^\mu\equiv\mathscr{K}^\mu \delta(p^2)$, the equation becomes
\begin{eqnarray}
\epsilon_{\mu\nu\rho\sigma} p^{\sigma} A^\rho
=-(p_\mu K_\nu - p_\nu K_\mu).\label{eq.Kconstrain}
\end{eqnarray}
Noting that the Vlasov equation as in Eq.(\ref{eq:041})
$$\delta(p^2)\, p^{\mu}\nabla_\mu f_\chi^{(0)}=0$$
requires $p_\mu A^\mu=0$, one can derive that
$$p_\mu K^\mu=0,\qquad A_\mu K^\mu=0.$$
It indicates that the unknown $K^\mu$ vector is orthogonal to two known vectors $A^\mu, p^\mu$ orthogonal to each other, the latter of which is a null-vector.
In principle, in the 3+1D space-time, there should be unique solution of $K^\mu$, with an undetermined component parallel to $p^\mu$.

To see this, let's first consider a simplified case: if taking the null-vector $p^\mu=(E,E,0,0)$, then one could always right down its orthogonal vectors as
$$A^\mu=(a,a,b,c), \qquad K^\mu=(k,k,d,f),$$
and $A,K$'s being orthogonal yields $bd+cf=0,$ which is similar to the 2D orthogonal condition.
Substituding this in Eq.(\ref{eq.Kconstrain}), one could find
$$d=-c,\quad f=b.$$
This indicates that for any given known $p^\mu$ and $A^\rho\equiv\frac{1}{2}\chi\Big(\nabla^\rho f_\chi^{(0)}\Big)\delta(p^2)$,
we can fix $K^\mu$ except its component parallel to $p^\mu$.
As a matter of fact, such conclusion is valid not only in the frame that $p^\mu=(E,E,0,0)$, but also in any general case.
Being any null vector, $p^\mu$ can always be express by its direction angle $\theta$ and $\phi$
\begin{eqnarray}
p^\mu = E(1, \sin\theta\cos\phi, \sin\theta\sin\phi, \cos\theta),
\end{eqnarray}
hence its two orthogonal vectors can be expanded in the corresponding basis:
\begin{eqnarray}
A^\mu &=& a(1, \sin\theta\cos\phi, \sin\theta\sin\phi, \cos\theta) + b(0, \cos\theta\cos\phi, \cos\theta\sin\phi, -\sin\theta) + c(0,-\sin\phi, \cos\phi,0),\\
K^\mu &=& k(1, \sin\theta\cos\phi, \sin\theta\sin\phi, \cos\theta) + d(0, \cos\theta\cos\phi, \cos\theta\sin\phi, -\sin\theta) + f(0,-\sin\phi, \cos\phi,0).\label{eq.Kcoef}
\end{eqnarray}

To solve $K^\mu$, we introduced an {\it arbitrary} auxiliary time-like vector $n^\mu=(n^t,n^x,n^y,n^z)$, normalized to unity: $n^\mu n_\mu=1$,
and construct the solution as in Eq.(\ref{eq.Kmu}),
\begin{eqnarray}
K^\mu &=& \frac{\epsilon^{\mu\nu\rho\sigma}p_\nu n_\rho A_\sigma}{n\cdot p}. \label{eq.Ksolution}
\end{eqnarray}
First of all, let's show that Eq.(\ref{eq.Ksolution}) gives a valid solution to Eq.(\ref{eq:069})/Eq.(\ref{eq.Kconstrain}). 
Substituting the solution to the right-hand-side of Eq.(\ref{eq.Kconstrain}), one obtains
\begin{eqnarray}
&&-(p_\mu K_\nu - p_\nu K_\mu) \nonumber\\
&=& \frac{p_\nu \epsilon_{\mu\alpha\rho\sigma}p^\alpha n^\rho A^\sigma}{n\cdot p} - \frac{p_\mu \epsilon_{\nu\alpha\rho\sigma}p^\alpha n^\rho A^\sigma}{n\cdot p} \\
&=& \frac{p_\nu \epsilon_{\mu\alpha\rho\sigma}p^\alpha n^\rho A^\sigma}{n\cdot p} 
+ \frac{p_\nu \epsilon_{\alpha\rho\sigma\mu}p^\alpha n^\rho A^\sigma}{n\cdot p}
+ \frac{p_\alpha \epsilon_{\rho\sigma\mu\nu}p^\alpha n^\rho A^\sigma}{n\cdot p}
+ \frac{p_\rho \epsilon_{\sigma\mu\nu\alpha}p^\alpha n^\rho A^\sigma}{n\cdot p}
+ \frac{p_\sigma \epsilon_{\mu\nu\alpha\rho}p^\alpha n^\rho A^\sigma}{n\cdot p} \\
&=& \frac{p_\nu \epsilon_{\mu\alpha\rho\sigma}p^\alpha n^\rho A^\sigma}{n\cdot p} 
- \frac{p_\nu \epsilon_{\mu\alpha\rho\sigma}p^\alpha n^\rho A^\sigma}{n\cdot p}
+ \frac{(p \cdot p) \epsilon_{\rho\sigma\mu\nu} n^\rho A^\sigma}{n\cdot p}
+ \frac{(n \cdot p) \epsilon_{\sigma\mu\nu\alpha}p^\alpha A^\sigma}{n\cdot p}
+ 0 \\
&=& \epsilon_{\mu\nu\rho\sigma} p^\sigma A^\rho,
\end{eqnarray}
which satisfies the equality.

Secondly, after some tedious but straightforward steps, one can compute the coefficients in Eq.(\ref{eq.Kcoef}) as
\begin{eqnarray}
d&=&-c,\label{eq.K1}\\
f&=&b,\\
k&=&\frac{b(-n^x\sin\phi + n^y\cos\phi) - c(n^x\cos\theta\cos\phi + n^y\cos\theta\sin\phi - n^z\sin\theta)}{p \cdot n}.\label{eq.Kparallel}
\end{eqnarray}
We can see explicitly that no matter what $n^\mu$ field we choose, it gives the same component of $K^\mu$ orthogonal to the momentum $p^\mu$.
It shows that Eq.(\ref{eq.Kmu}) gives a valid and complete solution of $\mathscr{K}^\mu$,
as long as we constrain $n^\mu$ to be time-like which ensures $p \cdot n\neq0$.

On the other hand, as can be seen in Eq.(\ref{eq.Kparallel}), different $n^\mu$ influence the component parallel to $p^\mu$.
To understand the role of $n^\mu$ and why it may cause ambiguity in $\mathscr{K}^\mu$, let's carefully consider the decomposition 
$\mathscr{H}^\mu\equiv p^\mu f^{(1)}_{\chi}+ \mathscr{K}^\mu$, trying to separate the vector $\mathscr{H}^\mu$ orthogonal to $p^\mu$ into two parts.
This decomposition is however subtle due to the light-like nature of $p$: $p^\mu p_\mu = 0$, i.e. $p$ is ``self-orthogonal''. 
It deserves commenting that this light-like feature is of course ultimately because the chiral fermion is massless.
To avoid ambiguity of the decomposition, one can always ensure that $f^{(1)}_\chi$ contains all $p^\mu$-parallel components by constraining 
$$p^0  \mathscr{K}^0 - \sum_{i=1}^{3}p^i \mathscr{K}^i = 0, \qquad p^0  \mathscr{K}^0 + \sum_{i=1}^{3}p^i \mathscr{K}^i = 0,$$
or equivalently, 
\begin{equation}
p\cdot\mathscr{K}=0, \qquad \mathscr{K}^0=0.
\end{equation}
Such requirement can be achieved by taking $n^\mu=(1,0,0,0)$, which yields $k=0$ in Eq.(\ref{eq.Kparallel}), and
\begin{eqnarray}
\mathscr{\overline{K}}^\mu = \frac{\epsilon^{\mu\nu\rho\sigma}p_\nu n_\rho (\nabla_\sigma f_\chi^{(0)})}{2\chi(n\cdot p)}\bigg|_{n=(1,0,0,0)} = \left(0 \;,\; -\frac{\chi}{2|\boldsymbol{p}|} \boldsymbol{p}\times(\boldsymbol{\nabla}f_{\chi}^{(0)})\right).\label{eq.Klab}
\end{eqnarray}
For an observer in the lab frame, Eq.(\ref{eq.Klab}) gives the complete decomposition of $\mathscr{H}^\mu$.
However, this is not the end of the story -- such characteristic is not boost-invariant, due to the fact that the requirement of ``orthogonality'' is not Lorentz-invariant. 
One can find a vector $\mathscr{K}^\mu$ orthogonal to a null-vector $p^\mu$ by restricting $\mathscr{K}^0=0,\;p \cdot \mathscr{K}=0$, but it's impossible to maintain
$\mathscr{K}'^0=0$ under arbitrary Lorentz transformation $\mathscr{K}'^\mu = \Lambda^{\mu}_{\;\,\nu}\mathscr{K}^\nu$, $p'^\mu = \Lambda^{\mu}_{\;\,\nu}p^\nu$.
To see this explicitly, for an observer moving with velocity $u^\mu$, the transformation $\Lambda^{\mu}_{\;\,\nu}$ is given by the element in the $(\mu+1)$-th row, $(\nu+1)$-th column of the matrix
\begin{eqnarray}
\left(
\begin{array}{cccc}
u^t & -u^x & -u^y & -u^z \\
-u^x & 1+\frac{u^x u^x}{1+u^t} & \frac{u^x u^y}{1+u^t} & \frac{u^x u^z}{1+u^t} \\
-u^y & \frac{u^y u^x}{1+u^t} & 1+\frac{u^y u^y}{1+u^t} & \frac{u^y u^z}{1+u^t} \\
-u^z & \frac{u^z u^x}{1+u^t} & \frac{u^z u^y}{1+u^t} & 1+\frac{u^z u^z}{1+u^t} \\
\end{array}
\right),
\end{eqnarray}
while in his local rest frame, 
\begin{eqnarray}
\mathscr{\overline{K}}'^0 
= \Lambda^{0}_{\;\,\nu} \mathscr{\overline{K}}^\nu = \frac{\chi}{2|\boldsymbol{p}|} \boldsymbol{u}\cdot\boldsymbol{p}\times(\boldsymbol{\nabla}f_{\chi}^{(0)}) 
= \frac{\chi}{2|\boldsymbol{p}|} (\boldsymbol{u}\times\boldsymbol{p})\cdot(\boldsymbol{\nabla}f_{\chi}^{(0)}) \neq 0.
\end{eqnarray}

Hence, the decomposition of $\mathscr{H}^\mu$ is frame dependent, and one should determine $f^{(1)}_{\chi}$ and $\mathscr{K}^\mu$ differently, with respect to different frame. As a matter of fact, for the observer moving with velocity $u^\mu$, one can construct $\mathscr{K}^\mu$ as
\begin{equation}
\mathscr{K}^\mu = \frac{\chi}{2u\cdot p}  \epsilon^{\mu\nu\rho\sigma} p_\nu u_\rho (\nabla_\sigma f^{(0)}_\chi),
\end{equation}
where the time-component of vector $\mathscr{K}$ vanishes in his local rest frame:
\begin{eqnarray}
\mathscr{K}'^0 = \Lambda^{0}_{\;\,\nu} \mathscr{K}^\nu
 = u_\nu \left(\frac{\chi}{2u\cdot p}  \epsilon^{\nu\mu\rho\sigma} p_\mu u_\rho (\nabla_\sigma f^{(0)}_\chi)\right)
 = \frac{\chi}{2u\cdot p} \epsilon^{\nu\mu\rho\sigma} p_\mu u_\nu u_\rho (\nabla_\sigma f^{(0)}_\chi)  = 0.
\end{eqnarray}
Actually, it's more obvious if one expresses all quantities in the observer's local rest frame:
\begin{eqnarray}
\mathscr{K}'^\mu 
&=&\Lambda^{\mu}_{\;\,\nu}\frac{\chi}{2u\cdot p}  \epsilon^{\nu\lambda\rho\sigma} p_\lambda u_\rho (\nabla_\sigma f^{(0)}_\chi)\nonumber\\
&=&\frac{\chi}{2u^{'}\cdot p^{'}}\Lambda^{\mu}_{\;\,\nu}\Lambda^{\;\,\nu}_{\alpha}\Lambda^{\;\,\lambda}_{\beta}\Lambda^{\;\,\rho}_{\kappa} \Lambda^{\;\,\sigma}_{\delta}\epsilon^{\alpha\beta\kappa\delta} p_\lambda u_\rho (\nabla_\sigma f^{(0)}_\chi)\nonumber\\
&=&\frac{\chi}{2u^{'}\cdot p^{'}}\epsilon^{\mu\beta\kappa\delta} p^{'}_{\beta} u^{'}_{\kappa} (\nabla^{'}_{\delta} f^{(0)}_\chi)\nonumber\\
&=& \frac{\chi}{2u'\cdot p'}  \epsilon^{\mu\nu\rho\sigma} p'_\nu u'_\rho (\nabla'_\sigma f^{(0)}_\chi)\bigg|_{u'=(1,0,0,0)}\nonumber \\
&=& \left(0 \;,\; -\frac{\chi}{2|\boldsymbol{p'}|} \boldsymbol{p'}\times(\boldsymbol{\nabla'}f_{\chi}^{(0)})\right).
\end{eqnarray}

%One can also note that when taking $n^\mu=(1,0,0,0)$ as in Sec.\ref{sec.3dCKT}, we can see explicitly that $k=0$.
%This implies that $\mathscr{K}^\mu$ is strictly orthogonal to $p^\mu$, and  Eq.(\ref{eq.Hmu})
%\begin{align}
%\mathscr{H}_{\mu}=p_{\mu}f^{(1)}_{\chi}+\mathscr{K}_{\mu}
%\end{align}
%separates, completely and strictly, the $\mathscr{H}^\mu$'s components parallel and orthogonal to the momentum $p^\mu$.

Consequently, one can see that constructing $\mathscr{K}_{\mu}$ as in Eq.(\ref{eq.Kmu}) with arbitrary time-like vector $n^\mu$ has the following physical meaning: 
for an observer moving with velocity $u^\mu=n^\mu$,
$\mathscr{K}^\mu \equiv \frac{\epsilon^{\mu\nu\rho\sigma}p_\nu n_\rho (\nabla_\sigma f_\chi^{(0)})}{2\chi(n\cdot p)}$
contains no $p^\mu$-parallel component in his local rest frame.
It gives a complete decomposition of $\mathscr{H}^\mu$, and $f^{(1)}_{\chi}$ corresponds to the first-order correction of the distribution function observed in this frame.
This reflects the frame dependence of spin tensor $S^{\mu\nu}\equiv \lambda\frac{\epsilon^{\mu\nu\rho\sigma}p_\rho n_\sigma}{p\cdot n}$ as mentioned in \cite{Hidaka:2016yjf,Hidaka:2017auj,Chen:2014cla}.

It's worth mentioning that $\mathscr{K}^\mu$ in Eq.(\ref{eq.Kmu}) is a vector defined in the lab frame, and once $n$ is fixed, it transforms like a Lorentz vector under boost transformation. It has the meaning of {\it what is known by an observer in the lab frame about the proper decomposition for another observer moving with velocity $n$}. 
As being illustrated in Eqs.(\ref{eq.K1}-\ref{eq.Kparallel}), the $\mathscr{K}^\mu$ vectors, corresponding to observers moving with velocities $u$ and $v$ respectively, differs with a $p^\mu$-parallel component:
\begin{eqnarray}
&&\mathscr{K}^\mu_{[u]} - \mathscr{K}^\mu_{[v]} \nonumber\\
&=& \frac{\chi}{2u\cdot p}  \epsilon^{\mu\nu\rho\sigma} p_\nu u_\rho (\nabla_\sigma f^{(0)}_\chi) 
- \frac{\chi}{2v\cdot p}  \epsilon^{\mu\nu\rho\sigma} p_\nu v_\rho (\nabla_\sigma f^{(0)}_\chi) \\
&=& \frac{\chi}{2(u\cdot p)(v \cdot p)} \left[ v_\alpha p^\alpha \epsilon^{\mu\nu\rho\sigma} p_\nu u_\rho (\nabla_\sigma f^{(0)}_\chi) 
- u_\alpha p^\alpha \epsilon^{\mu\nu\rho\sigma} p_\nu v_\rho (\nabla_\sigma f^{(0)}_\chi)\right] \\
&=& \frac{\chi}{2(u\cdot p)(v \cdot p)} \Bigg[ v_\alpha p^\alpha \epsilon^{\mu\nu\rho\sigma} p_\nu u_\rho (\nabla_\sigma f^{(0)}_\chi) 
+ u_\alpha p^\mu \epsilon^{\nu\rho\sigma\alpha} p_\nu v_\rho (\nabla_\sigma f^{(0)}_\chi)
+ u_\alpha p^\nu \epsilon^{\rho\sigma\alpha\mu} p_\nu v_\rho (\nabla_\sigma f^{(0)}_\chi) \nonumber\\
&&\qquad\qquad\qquad + u_\alpha p^\rho \epsilon^{\sigma\alpha\mu\nu} p_\nu v_\rho (\nabla_\sigma f^{(0)}_\chi)
+ u_\alpha p^\sigma \epsilon^{\alpha\mu\nu\rho} p_\nu v_\rho (\nabla_\sigma f^{(0)}_\chi) \\
&=& \frac{\chi}{2(u\cdot p)(v \cdot p)} \Bigg[ v_\alpha p^\alpha \epsilon^{\mu\nu\rho\sigma} p_\nu u_\rho (\nabla_\sigma f^{(0)}_\chi) 
+ p^\mu \epsilon^{\nu\alpha\rho\sigma} p_\nu u_\alpha v_\rho (\nabla_\sigma f^{(0)}_\chi)
+ (p \cdot p) \epsilon^{\mu\rho\alpha\sigma}v_\rho u_\alpha  (\nabla_\sigma f^{(0)}_\chi) \nonumber\\
&&\qquad\qquad\qquad - v_\rho p^\rho \epsilon^{\mu\nu\alpha\sigma} p_\nu u_\alpha  (\nabla_\sigma f^{(0)}_\chi)
+  \epsilon^{\mu\nu\alpha\rho}p_\nu u_\alpha v_\rho (p \cdot \nabla f^{(0)}_\chi)
\Bigg] \\
&=& \frac{\chi \epsilon^{\nu\alpha\rho\sigma} p_\nu u_\alpha v_\rho (\nabla_\sigma f^{(0)}_\chi)}{2(u\cdot p)(v \cdot p)} p^\mu.
\end{eqnarray}
Noting that the vector $\mathscr{H}^\mu$ should be frame independent
\begin{eqnarray}
 \mathscr{H}^\mu_{[u]} = \mathscr{K}^\mu_{[u]} + p^\mu f^{(1)}_{[u],\chi}
\quad&\equiv&\quad \mathscr{H}^\mu_{[v]} = \mathscr{K}^\mu_{[v]} + p^\mu f^{(1)}_{[v],\chi},
\end{eqnarray}
one can see explicitly the difference between distributions observed in $u$- and $v$-frames:
\begin{eqnarray}
 f_{[u],\chi} -  f_{[v],\chi} = \hbar (f^{(1)}_{[u],\chi} -  f^{(1)}_{[v],\chi}) = -\frac{\hbar \chi \epsilon^{\nu\alpha\rho\sigma} p_\nu u_\alpha v_\rho (\nabla_\sigma f^{(0)}_\chi)}{2(u\cdot p)(v \cdot p)}.
\end{eqnarray}

\end{appendix}

%\end{CJK}
\end{document}